\documentclass{article}

\usepackage[english]{babel}

\usepackage[letterpaper,top=2cm,bottom=2cm,left=3cm,right=3cm,marginparwidth=1.75cm]{geometry}

\usepackage{amsmath}
\usepackage{graphicx}
\usepackage{xcolor}
\usepackage{amssymb}
\usepackage{calc}
\usepackage{mathtools}
\usepackage{caption}
\usepackage{subcaption}
\usepackage{comment}
\usepackage{lineno}
\usepackage{slashed}
\usepackage[normalem]{ulem}
\usepackage[english]{babel}
\usepackage[colorlinks=true, allcolors=blue]{hyperref}

\begin{document}
\Large
\begin{center}
{\bf  The Finite Geometry of Breaking Quantum Secrets}
\end{center}
\large
\vspace*{-.1cm}
\begin{center}
P\'eter L\'evay$^{1,2}$
and Metod Saniga$^{3}$ \end{center}
 \vspace*{-.4cm} \normalsize
 \begin{center}
$^{1}$ 
HUN-REN-BME-BCE Quantum Technology Research Group\\
Budapest University of Technology and Economics\\
Műegyetem rkp. 3., H-1111 Budapest, Hungary
\end{center}
\begin{center}
$^{2}$ 
Department of Theoretical Physics\\
Institute of Physics\\
Budapest University of Technology and Economics\\
Műegyetem rkp. 3., H-1111 Budapest, Hungary
\end{center}
\begin{center}
$^{3}$
 Astronomical Institute, Slovak Academy of Sciences\\ SK-05960 Tatranská Lomnica, Slovak Republic 

\vspace*{.0cm}
\vspace*{.2cm} (13 February 2026)
\end{center}
\vspace*{-.3cm} \noindent \hrulefill

\vspace*{.3cm}
  \noindent\\

\begin{abstract}

Using a finite geometric framework for studying the pentagon and
heptagon codes we show
that the concepts of quantum secret sharing and contextuality can be
studied in a nice and
unified manner. The basic idea is a careful study of the respective 2 +
3 and 3 + 4
tensorial factorizations of the elements of the stabilizer groups of
these codes. It is
demonstrated in detail how finite geometric structures entailing a
specific three-qubit
(resp. four-qubit) embedding of binary symplectic polar spaces of rank
two (resp. three),
corresponding to these factorizations, govern issues of contextuality
and entanglement
needed for a geometric understanding of quantum secret sharing.
Using these results for the $(3,5)$ and $(4,7)$ threshold schemes explicit secret breaking protocols are derived.
Our results hint at a novel geometric way of looking at contextual configurations.

\end{abstract}

\section{Introduction}

This work is about the finite geometry 
of the pentagon\cite{Bennett,Laflamme} and heptagon\cite{Steane} codes.
Both codes can be regarded as the archetypical examples of encoding structures needed for building up holographic codes\cite{Happy,Holohepta}. As is well-known such codes can successfully be used for understanding discretized models of emergent spacetime structures\cite{Raamsdonk,Almheiri}.
These are
 geometrical ones which are coming from quantum entanglement via the dualities built in the framework of the AdS/CFT correspondence\cite{Maldacena}.
In a recent paper one of us have argued\cite{LevayToy} that the basic building blocks of such encoding structures could be finite geometric ones. The conjecture of that paper was that these structures can serve as building blocks  encapsulating  all the fundamental quantum notions needed for our understanding of spacetime as a quantum error correcting code\cite{Almheiri}.

In this paper however, instead of elaborating further on these exciting ideas, we are content with a simple demonstration of an intriguing connection between finite geometry and quantum secret sharing schemes. Such a result could be interesting for the quantum information community and also for researchers dealing with fundamental issues of quantum theory. In particular our goal is to point out how finite geometric structures show up in the codes underlying these schemes in concert with  basic quantum notions like entanglement\cite{Schrodinger,EPR}, contextuality\cite{Bell,KochenSpecker,Mermin}, quantum nets\cite{GHW} and mutually unbiased bases\cite{MUB}.

The elaborations of our paper are based on the well-known stabilizer formalism\cite{Gottesman,NielsenChuang}. In the case of the pentagon (heptagon) code our simple strategy is to look at the explicit structure of the $15$ ($63$) nontrivial observables
forming the Abelian stabilizer group of the code.
The penta (hepta) code is encoding one logical qubit into $5$ ($7$) qubits.
In the case of the penta (hepta) code these observables can be
represented by $10$ ($14$) component vectors with elements taken from the field $GF(2):={\mathbb Z}_2\equiv \{0,1\}$.
In this representation the commutation properties of the observables are taken care by a symplectic form on these vector spaces\cite{Calderbank,NielsenChuang}. This is a $GF(2)$-valued nondegenerate bilinear form taking the value $0$ for commuting and the value $1$ for anticommuting observables. 

For our finite geometric considerations it is rewarding to use a projective geometric setup. Then the presence of the symplectic form even in the projective context lends itself to the possibility of defining finite incidence geometric structures in these spaces.
Informally: two points are incident if the corresponding observables are commuting and non incident if they are anti commuting\cite{SP}.
In this setup the nontrivial elements of stabilizer groups then show up as points living in the projective spaces $PG(9,2)$ ($PG(13,2)$). 
In particular sets of mutually commuting observables of our codes turn out to be represented by points belonging to $3$ ($5$)
dimensional totally isotropic subspaces of  $PG(9,2)$ ($PG(13,2)$).
Taking this condition of "total isotropy" representing "mutually commuting" into account we arrive at our main actors on the scene: the symplectic polar spaces $W(9,2)$ ($W(13,2)$).
These are mathematical objects containing totally isotropic subspaces (with their maximal dimensions being $4$ ($6$)) encapsulating the commutation properties of sets of observables in an incidence geometrical manner.

This formalism then will be used to arrive at a geometric understanding of the $(3,5)$ ($(4,7)$) secret sharing schemes\cite{How}.
Under these schemes, at least $3$ ($4$) parties must cooperate in order to break the secret. 
Then for the pentagon (heptagon) code we invoke a $2+3$ ($3+4$) split. We call parties of cardinality $n=2$ ($n=3$) as the ones who are not willing to cooperate, and the ones of cardinality $k=3$ ($k=4$) as the ones who are willing to cooperate in breaking the secret.
Under this split then we obtain {\it two} families of $15$ ($63$) observables. Now unlike the original {\it commuting} set of observables representing totally isotropic subspaces, both of the split families of observables contain commuting and anticommuting elements as well. Due to this fact the non cooperating parts of the families can be used to
parametrize the nontrivial incidence geometries of the symplectic polar spaces  $W(3,2)$ ($W(5,2)$).
On the other hand the cooperating parts give rise to incidence structures of the same kind this time in a representation coming from the embeddings of these spaces to the ones $W(5,2)$ ($W(7,2)$).

The basic object underlying our considerations is the space of contexts. In the case of the pentacode (heptacode) these are triples (heptads) of mutually commuting observables.
They correspond to the lines (planes) of the space $W(3,2)$ ($W(5,2)$).
It turns out that the space of contexts contains $15$ ($135$) points representing the $15$ ($135$) totally isotropic lines (planes) of $PG(3,2)$ ($PG(5,2)$).
As is well-known from the theory of contextuality\cite{Bell,KochenSpecker}
observables can form positive or negative lines that can be used for proving Kochen-Specker-like theorems\cite{Mermin}.
It turns out that one can generalize the concept of positivity (negativity) for higher dimensional subspaces too.

Then in the $n=2$ ($n=3$) case $3$ ($9$) negative lines (planes) labeled by two-qubit (three-qubit) observables
play an important role. Moreover, as a result of our split and the fact that
$k > n$  also gives rise to nontrivial embeddings, 
in our considerations such negative lines (planes) labeled with three-qubit, (four-qubit) observables
also appear. This upgrades our considerations with the presence of special embedded negative lines (planes)
$k=3$ ($k=4$) displaying interesting geometric structure.

Apart from their fascinating structure, they are also useful.
We show that by elevating these subspaces to positive ones in different ways makes us possible to associate to
these special points of context space different branches of biseparable states as components of the stabilized states. The presence of these
biseparable branches in the pentagon (heptagon) stabilizer state makes it possible for one of the parties to possess the secret. We will give the different decompositions explicitly in an elegant geometric manner. They are displaying the protocols needed to be performed
by the cooperating parties enabling one of them to succeed in achieving her/his goal.  
We must stress at this point that similar decompositions have already been presented in the literature (see for instance\cite{China}) however, their underlying geometry according to our best knowledge have never been explored and connected to other quantum notions.
Our paper hopefully remedy this situation.

The organization of this paper is as follows.
Section 2. is devoted to the pentagon and Section 3. to the heptagon code.
In different subsections of these sections we start by recalling basic well-known facts of the code, then immediately proceed to analyzing the relevant finite geometric structures.
Then issues of contextuality and the entanglement structure of the corresponding states are briefly summarized using the stabilizer formalism. Here notions like negative lines and planes show up. 
Then we proceed to finding finite geometry based representations of the corresponding stabilizer states.
We explicitly write down the geometric decompositions of these states, and analyze the protocols needed for secret breaking.
Section \ref{sec:conc} is devoted to the conclusions and  comments. Here, in Section \ref{sec:comment} a brief reminder is also included to point out how our results can be connected to a recent finite geometric model of discretized space time functioning as an error correcting code\cite{LevayToy}.
This comment gives a hint on how our results might fit into the larger picture mentioned in the beginning of this introduction.

We tried to keep the main body of our paper at a short and elementary level, devoid of mathematical details, stressing physical intuition instead.  
However, for the convenience of a reader who is in need of more mathematical details and subtlety
the precise finite geometric definitions, and other more sophisticated explanations are relegated to an  Appendix. This material containing detailed elaborations can be found in the subsections of Section 6.

\bigskip

\section{The Pentagon Code}

\subsection{Stabilizer formalism}

The pentagon code\cite{Laflamme,Bennett} is encoding one logical qubit into five qubits.
In the language of stabilizer codes\cite{Gottesman,Calderbank} it works as follows.
First we take the Abelian group (stabilizer group) ${\mathcal S}$
in its generator presentation of the form
\begin{equation}
{\mathcal S}=\langle g_1,g_2,g_3,g_4\rangle \equiv \langle {\color {red} XZ}ZXI,{\color {red}IX}ZZX,{\color {red} XI}XZZ,{\color {red}ZX}IXZ\rangle,
\label{presentation1}
\end{equation}
where $X,Y,Z$ are the Pauli operators and $I$ is the identity operator represented by the usual $2\times 2$ matrices.
They act in the usual manner
in the computational basis
\begin{equation}
X\vert 0\rangle =\vert1 \rangle,\qquad X\vert 1\rangle =\vert 0\rangle  \nonumber
\end{equation}
\begin{equation}
Y\vert 0\rangle =i\vert1 \rangle,\qquad Y\vert 1\rangle =-i\vert 0\rangle 
\nonumber
\end{equation}
\begin{equation}
Z\vert 0\rangle =\vert 0 \rangle,\qquad Y\vert 1\rangle =-\vert 1\rangle.  
\nonumber
\end{equation}
\noindent
In Eq.(\ref{presentation1}) tensor products are omitted,
i.e. $XZZXI$ is a shorthand for the observable
$X\otimes Z\otimes Z\otimes X\otimes I$.
The color and the $2+3$ split will be explained soon.
The generators in the (\ref{presentation1}) presentation are commuting 
hence ${\mathcal S}$ is a group of order $16$.

For our finite geometric considerations we need all the elements of this group hence we also give here a list.
We have
\begin{equation}
\{g_1g_2, g_1g_3, g_1g_4, g_2g_3, g_2g_4, g_3g_4\}=
\{{\color {red}XY}IYX,{\color {red}IZ}YYZ,{\color {red} YY}ZIZ,{\color {red}XX}YIY,{\color {red}ZI}ZYY,{\color {red} YX}XYI\},
\label{presentation2}
\end{equation}

\begin{equation}
\{g_2g_3g_4, g_1g_3g_4, g_1g_2g_4, g_1g_2g_3, g_1g_2g_3g_4\}=
\{{\color {red}YI}YXX,{\color {red}ZY}YZI,{\color {red} YZ}IZY,{\color {red}IY}XXY,{\color {red}ZZ}XIX\},
\label{presentation3}
\end{equation}
taken together with the identity ${\bf 1}\equiv{\color {red} II}III$ we have all the $16$ elements.

The easiest way to remember all the elements of this Abelian group is to notice its permutation symmetry\cite{Peres}. If we take the cyclic permutations of the five operators featuring the three distinguished elements $XYIYX$,
$YZIZY$ and $ZXIXZ$ one can generate all the $15$ nontrivial elements of ${\mathcal S}$.
Indeed,
\begin{equation}
\{XYIYX,XXYIY,YXXYI,IYXXY,YIYXX\}=\{g_1g_2,g_2g_3,g_3g_4,g_1g_2g_3,g_2g_3g_4\},
\nonumber
\end{equation}

\begin{equation}
\{YZIZY,YYZIZ,ZYYZI,IZYYZ,ZIZYY\}=\{g_1g_2g_4,g_1g_4,g_1g_3g_4,g_1g_3,g_2g_4\},
\nonumber
\end{equation}

\begin{equation}
\{ZXIXZ,ZZXIX,XZZXI,IXZZX,XIXZZ\}=\{g_4,g_1g_2g_3g_4, g_1,g_2,g_3\}.
\nonumber
\end{equation}
In the following we will be interested in $2+3$ splits of these $15$ operators. Due to the permutation symmetry one can consider any of the five qubits playing a special role with respect to the split. Here we have chosen the first and second qubits to be special. These are the red ones appearing in our first list of Eqs.(\ref{presentation1})-(\ref{presentation3}).

With this split we are intending to understand the finite geometric aspects of a quantum secret sharing scheme\cite{How} of $(3,5)$ type.
As it is well-known under this scheme a quantum secret is divided into $5$ shares (qubits) such that any $3$ of those
shares (qubits) can be used to reconstruct the secret, but any set of
fewer shares (qubits) contains absolutely no information
about the secret. Generally such schemes are called  $(k, m)$ threshold schemes\cite{How}.

Back to the theory of stabilizer codes we have $m=5$ qubits and $l=4$ independent generators hence the observables of $\mathcal S$ stabilize a $2^{m-l}=2$ dimensional subspace of the five qubit space\cite{Gottesman,NielsenChuang}. This is the subspace of the logical qubit encoded into the five qubit Hilbert space.

In order to see explicitly this subspace, in the $32$ dimensional vector space of five qubits, we use the projectors $P_j=\frac{1}{2}({\bf 1}+g_j), j=1,2,3,4$ to generate it 
from two linearly independent five qubit states. 
Notice first that the operator 
$P_4P_3P_2P_1$
contains all of the $16$ operators of the group listed above. Then applying this operator
to the states $\vert00000\rangle$
and $\vert 11111\rangle $ one gets the basis vectors $\vert\bar{0}\rangle$
and
$\vert\bar{1}\rangle$.
These are then spanning the stabilized subspace where our logical qubit resides.
Indeed, by construction the states of the form
\begin{equation}
\vert\Psi\rangle=\alpha\vert\bar{0}\rangle +\beta\vert\bar{1}\rangle
\label{psi}
\end{equation}
are stabilized by all of the elements of ${\mathcal S}$, i.e. we have the property
$g\vert\Psi
\rangle=\vert\Psi\rangle,\quad \forall g\in {\mathcal S}, \quad\alpha,\beta\in \mathbb{C}$.

A calculation shows that for $\vert\bar{0}
\rangle =NP_4P_3P_2P_1\vert 00000\rangle$ 
and
$\vert\bar{1}
\rangle =NP_4P_3P_2P_1\vert 11111\rangle =XXXXX\vert\bar{0}\rangle$
with the normalization factor $N=4$ we have the explicit forms

\begin{equation}
\begin{split}
4\vert\bar{0}\rangle=\vert 00000\rangle
&+\vert 10100\rangle+\vert 01010\rangle+\vert00101\rangle
+\vert 10010\rangle
+\vert 01001\rangle\\
&-\vert 11000\rangle
 -\vert 01100\rangle
 -\vert 00110\rangle
 -\vert 00011\rangle
 -\vert 10001\rangle\\
&-\vert 01111\rangle
 -\vert 10111\rangle
 -\vert 11011\rangle
 -\vert 11101\rangle
 -\vert 11110\rangle,
\end{split}
\nonumber
\end{equation}

\begin{equation}
\begin{split}
4\vert\bar{1}\rangle=\vert
11111\rangle
&+\vert 01011\rangle+\vert 10101\rangle+\vert 11010\rangle
+\vert 01101\rangle
+\vert 10110\rangle\\
&-\vert 00111\rangle
 -\vert 10011\rangle
 -\vert 11001\rangle
 -\vert 11100\rangle
 -\vert 01110\rangle\\
&-\vert 10000\rangle
 -\vert 01000\rangle
 -\vert 00100\rangle
 -\vert 00010\rangle
 -\vert 00001\rangle.
\nonumber
\end{split}
\end{equation}
Up to conventions these are the states that can be found in the equivalent formulations of our code in the classical papers of Refs.\cite{Bennett,Laflamme,Peres}.

\subsection{The Doily}

Let us now consider our $2+3$ split of the $15$ nontrivial observables comprising $\mathcal S$.
Associate points to these two qubit or three qubit   observables. Then
connect the corresponding triples of observables that are mutually commuting.
What we get is a finite geometric object called the "doily". The precise mathematical definition of this object is $W(3,2)$ which is an abbreviation for the term : a symplectic polar space of rank two and order two\cite{SP}. 
For the definitions see Appendix \ref{sec:app3}.

Now thanks to our split we have two different labelings of the points of the doily. One of them with two qubit observables and the other with three qubit ones. For the two qubit case {\it all of the two qubit observables} can be mapped bijectively to nontrivial doily points. On the other hand {\it some of the three qubit observables} are mapped to doily points.
Which $15$ out of the $63$ nontrivial three-qubit observables shows up in the three-qubit labeling is dictated by the structure of the pentagon code.
Hence the pentagon code for an arbitrary split provides an embedding
of two qubit observables 
 to certain three-qubit ones.

\begin{figure}
\centering
\includegraphics[width=0.45\linewidth]{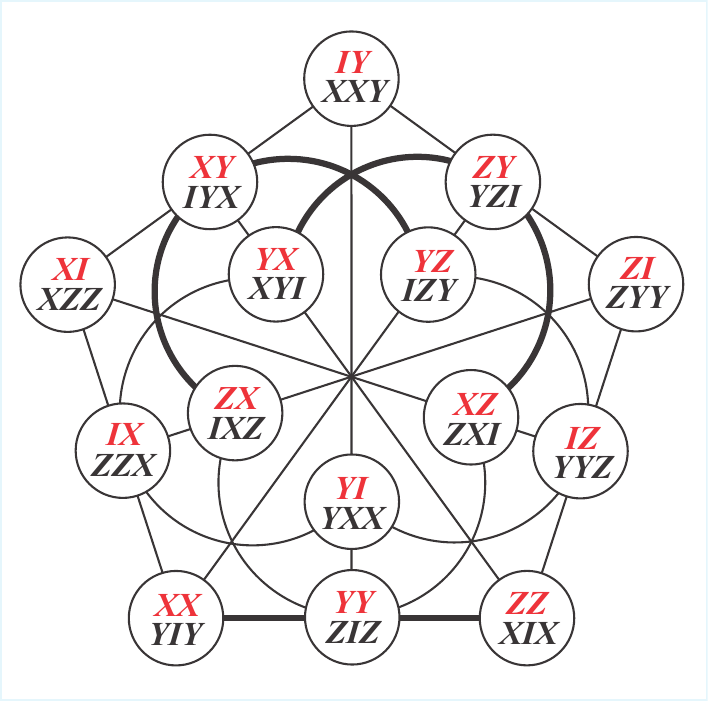}
\caption{The doily with the split labeling.
This means that the nontrivial five qubit observables of ${\mathcal S}$ of (\ref{presentation1})-(\ref{presentation3}) are split into two.
One label is a two qubit one and the other is a three qubit one.
Here the two qubit labels are corresponding to the red "digits" and the three qubits ones to the remaining ones of our list of the $15$ nontrivial elements of the group ${\mathcal S}$.
Observables connected by a "line" are mutually commuting both for the two and three qubit labelings. 
The bold faced lines are the negative lines. All other lines are positive ones.
Notice that the doily structure is emerging only after invoking the $2+3$ split.
Without this split all of the $5$ qubit observables are mutually commuting. 
}
\label{fig:doily}
\end{figure}

A pictorial representation of the doily with its two different types of labelings can be seen in Figure~\ref{fig:doily}. 
Notice  that in Figure~\ref{fig:doily}  with either type of  labeling among the $15$  observables we have both commuting and anticommuting ones (the latter ones are not connected by lines).
This is to be contrasted with the unsplitted $15$ five qubit operators of the Abelian $\mathcal S$ with all of the observables mutually commuting.

From the finite geometric point of view we have the following situation.
The five qubit Pauli group modulo its center is a $10$ dimensional vector space over $GF(2):={\mathbb Z}_2\equiv
\{0,1\}$ denoted by $V(10,2)$.
This vector space is equipped with a symplectic form. It is a map from $V(10,2)$ to $GF(2)$.
The symplectic product of two vectors is zero when the corresponding Pauli operators are commuting and one if they are not commuting. 
$\mathcal S$ can be regarded as a maximal totally isotropic subspace of the projective space $PG(9,2)$ with respect to this symplectic form.
The projective dimension of $\mathcal S$ is $3$.
The two qubit doily is the symplectic polar space $W(3,2)$ consisting of the points of $PG(3,2)$ ($15$ points)
and its maximal totally isotropic lines ($15$ lines).
For more details and the precise definitions of these finite geometric structures see Appendix \ref{sec:app3}.

\subsection{Contextuality}

Observe that multiplying the operators along the lines of Figure~\ref{fig:doily} we get either $\pm II$ or $\pm III$ depending on whether we use the two or three qubit labeling for our doily.
Depending on the sign we call these lines either positive or negative lines.
In the figure we have three negative lines. They are highlighted as bold faced
lines. These are as follows:
\begin{equation}
\{{\color {red}XY},{\color {red}YY},{\color {red}ZZ}\},\quad \{{\color {red}ZX},{\color {red}XY},{\color {red}ZY}\},\quad  \{{\color {red}YX},{\color {red}ZY},{\color {red}XZ}\}
\end{equation}
for the two qubit part and
\begin{equation}
\{YIY,ZIZ,XIX\},\quad \{IXZ,IYX,IZY\},\quad \{XYI,YZI,ZXI\}
\end{equation}
for the three qubit part.
The other $12$ lines in both the two and three qubit cases are positive.
Notice that without the $2+3$ split for the $5$ qubit observables the same triples all yield "positive lines".
Hence the presence of both type of lines is due to the split of the system into two subsystems.

Now using the stabilizer formalism one can also realize
that the positive two qubit lines determine states and the three qubit positive ones
two dimensional subspaces. For a detailed example on this point see Appendix \ref{sec:app1}.

However, one can easily make the negative lines
to positive ones by flipping one of the signs. 
For example flipping the sign of $XZ$ is turning the set
$\{II,XZ,ZY,YX\}$ to a genuine stabilizer group,
call it ${\mathcal S}_A$, with its stabilized state
\begin{equation}
{\mathcal S}_A:=\langle -XZ,ZY\rangle,\quad
    \vert A\rangle=\frac{1}{2}(\vert 00\rangle-\vert 10\rangle+i\vert 11\rangle +i\vert 01\rangle ).
\end{equation}  
\noindent
Similary, via sign flips we have the pairs
\begin{equation}
{\mathcal S}_B:=\langle -XX,ZZ\rangle,\qquad
    \vert B\rangle=\frac{1}{\sqrt{2}}(\vert 00\rangle-\vert 11\rangle),
\end{equation}
\begin{equation}
{\mathcal S}_C:=\langle -ZX,YZ\rangle,\qquad
    \vert C\rangle=\frac{1}{2}(\vert 00\rangle-\vert 01\rangle +i\vert 11\rangle +i\vert 10\rangle).
\end{equation}
\noindent
Notice that there are four different stabilizer groups that can be associated with each line of the two qubit doily. For example one has
\begin{equation}
{\mathcal S}^{\pm\pm}:=\langle
\pm XX,\pm ZZ\rangle,    
\end{equation}
with the corresponding four stabilizer states being the four Bell states of Eqs.(\ref{Bell1})-(\ref{Bell4}) of Appendix \ref{sec:app1}. For example for ${\mathcal S}^{-+}\equiv {\mathcal S}_B$ we have the stabilized Bell state $\vert\varphi^{-+}\rangle\equiv\vert B\rangle $ and
the stabilizer state corresponding to 
${\mathcal S}^{+-}$ is $\vert\varphi^{+-}\rangle$ etc.

It is known that the doily, aka $W(3,2)$, is contextual with its degree of
contextuality being three, not only in
its basic two-qubit labeling\cite{degree} but also when located in its
multi-qubit generalization i.e. the space $W(2N-1,2)$, for any
any number of qubits $N>2$ (for the definition of this space see Appendix \ref{sec:app3}, see also Proposition 1 of Ref.\cite{degree} and Ref.\cite{degree2}).
This means that it is impossible
to assign signs to the points of the doily such that it is compatible with the assignment of observables producing either positive or negative lines.

Alternatively one can say that the doily is contextual because it is impossible to associate to all of its lines stabilizer groups by flipping the signs of some observables consistently.
The minimum number of lines obstructing this assignment of stabilizer groups to them defines the degree of contextuality. 
This point is further clarified in Appendix \ref{sec:appdoily}.
For more details on the literature of contextuality\cite{Bell,KochenSpecker,Mermin} see Refs.\cite{Cabello,degree,sheaf} and references therein.

\subsection{Breaking the secret with the negative lines of the doily}

By secret we mean a single qubit state
\begin{equation}
\vert\psi\rangle :=\alpha\vert 0\rangle+\beta\vert 1\rangle,\qquad \alpha,\beta\in{\mathbb C}
\label{secret}
\end{equation}
possessed by someone, say Alice.
Alice then encrypts her secret into
five qubits using the  pentagon code. After the encryption she has the state
$\vert\Psi\rangle =\alpha\vert\bar{0}\rangle +\beta\vert\bar{1}\rangle$
of Eq.(\ref{psi}).
Then $\vert\Psi\rangle$
is given to five parties, each of them is possessing merely one of the qubits of the corresponding subsystems, in the hope that the secret is in good hands. 

However, it is easy to see that any of the negative lines of the doily contains enough information for at least three cooperating parties
to perform suitable joint manipulations on their shares resulting in the breaking of the secret.
The breaking of the secret means that after the collaboration one of the parties (out of the three) succeeds in transforming the state of her/his share to $\vert\psi\rangle$.

First of all due to permutation symmetry any of the $2+3$ splits will do. Hence it is enough to consider the situation of subsystems $345$ cooperating for breaking the secret. This choice yields the split displayed in the set of observables of Eqs.(\ref{presentation1})-(\ref{presentation3}).
Now in order to see  how one of the negative lines of the doily helps for breaking the secret choose the line of Figure~\ref{fig:doily} labeled as $(YIY,ZIZ,XIX)$. 
By flipping a suitable number of signs this line can be converted to a positive one in four different ways. One of them is $(YIY,ZIZ,-XIX)$.
Notice then that for this choice for $\vert\Psi\rangle$ given by Eq. (\ref{psi}) we have
\begin{equation}
    ({\bf 1}-IIXIX)({\bf 1}+IIZIZ)\vert \Psi\rangle
    =(\alpha{\bf 1}+\beta XXXXX)(\vert 01\rangle +\vert 10\rangle)_{12}\otimes (\vert 010\rangle -\vert 111\rangle )_{345}.
\end{equation}
One can then notice that the resulting state in the $345$ part is a biseparable three-qubit state. The split is of the form: $(35)(4)$.
This can also be  seen from the biseparable structure of the stabilizer group $S^{+-}$ or the stabilized subspace $\vert\Phi^{+-}\rangle$ given by Eqs. (\ref{pm}) and (\ref{fipm}).
Alternatively with the Bell states of Eqs.(\ref{Bell1})-(\ref{Bell4}) and the projectors
\begin{equation}
\left(P_{\pm}\right)_{35}:=\frac{1}{2}({\bf 1}\pm IIXIX),\qquad
\left(Q_{\pm}\right)_{35}:=\frac{1}{2}({\bf 1}\pm IIZIZ)
\label{confuse}
\end{equation}
our equation can be written as
\begin{equation}
    2\left(P_-\right)_{35} \left(Q_{+}\right)_{35}\vert \Psi\rangle
    =\vert \varphi^{+-}\rangle_{12}\otimes\vert\varphi^{-+}\rangle_{35}\otimes (\alpha \vert 1\rangle -\beta \vert 0\rangle )_4.
    \label{proto1}
\end{equation}
Proceeding similarly with the other combinations of projectors, corresponding to the three remaining possibilities of converting our negative line to a positive one, finally we have
\begin{equation}
2\left(P_-\right)_{35} \left(Q_{+}\right)_{35}\vert\Psi\rangle =
\vert \varphi^{+-}\rangle_{12}\otimes\vert\varphi^{-+}\rangle_{35}\otimes XZ\vert\psi\rangle_4,
\label{finalresult1}
\end{equation}
\begin{equation}
2\left(P_-\right)_{35} \left(Q_{-}\right)_{35}\vert\Psi\rangle =
\vert \varphi^{--}\rangle_{12}\otimes\vert\varphi^{--}\rangle_{35}\otimes \vert\psi\rangle_4,
\label{finalresult2}
\end{equation}
\begin{equation}
2\left(P_+\right)_{35} \left(Q_{-}\right)_{35}\vert\Psi\rangle =
\vert \varphi^{++}\rangle_{12}\otimes\vert\varphi^{+-}\rangle_{35}\otimes (-X)\vert\psi\rangle_4,
\label{finalresult3}
\end{equation}
\begin{equation}
2\left(P_+\right)_{35} \left(Q_{+}\right)_{35}\vert\Psi\rangle =
\vert \varphi^{-+}\rangle_{12}\otimes\vert\varphi^{++}\rangle_{35}\otimes Z\vert\psi\rangle_4.
\label{finalresult4}
\end{equation}
Summing these states one obtains
\begin{equation}
\label{branching}
\begin{split}
2\vert\Psi\rangle=
&\vert\varphi^{--}\rangle_{12}\otimes\vert\varphi^{--}\rangle_{35}\otimes\vert\psi\rangle+
\vert\varphi^{-+}\rangle_{12}\otimes\vert\varphi^{++}\rangle_{35}\otimes Z\vert\psi\rangle +
\\
&\vert\varphi^{+-}\rangle_{12}\otimes\vert\varphi^{-+}\rangle_{35}\otimes XZ\vert\psi\rangle-
\vert\varphi^{++}\rangle_{12}\otimes\vert\varphi^{+-}\rangle_{35}\otimes X\vert\psi\rangle.
\end{split}
\end{equation}
Now from this form one can see that in order to retrieve the secret by the $4$th party, parties $345$ should proceed through similar steps than the ones familiar from the famous teleportation protocol\cite{Teleportation}. 

Namely, 
let us suppose that all of the parties associated with subsystems $345$ are aware of the (\ref{branching}) decomposition. Then
we see that if the parties of $35$ perform a von Neumann measurement in the Bell basis on their combined subsystem then their measurement outcome will reveal which of the four component states $\vert\varphi^{\pm\pm}\rangle_{35}$ is realized with probability $1/4$. 
Using then a classical communication channel, parties $35$ can communicate their result to party $4$.
Depending on this information $4$ either does nothing (see first term in the right hand side of Eq.(\ref{branching})) or performs either of the following set of unitary transformations on her/his share: $Z, XZ, -X$.
Now thanks to these operations the secret $\vert\psi\rangle$
can be reproduced as the state of the $4$th subsystem.

Let us now consider another split triple of negative lines of the doily namely (${\color {red}YX},{\color {red}ZY},{\color {red} XZ}$), ($XYI,YZI,ZXI$).
Using these triples one can see that the secret can be reproduced at the $5$th qubit as follows.
Let us first introduce the SWAP operator ${\bf S}$ swapping the first and second qubits.
Moreover, in Appendix \ref{sec:app1}. the four possible stabilizer groups associated with the corresponding negative lines are also written down.
For the corresponding four Bell-like states $\vert\chi^{\pm\pm}\rangle$ see Eqs.(\ref{bell1})-(\ref{bell4}).
Let us also introduce the single qubit unitary operator $U$ of Eq.(\ref{Yunitary}).
Then with the help of these quantities one can show that
\begin{equation}
2\left({\mathcal P}_+{\mathcal Q}_{+}\right)_{34}\vert\Psi\rangle =
{\bf S}\vert \chi^{++}\rangle_{12}\otimes\vert\chi^{++}\rangle_{34}\otimes e^{i\pi/4}UZ\vert\psi\rangle_5
\label{Finalresult1}
\end{equation}
\begin{equation}
2\left({\mathcal P}_-{\mathcal Q}_{-}\right)_{34}\vert\Psi\rangle =
{\bf S}\vert \chi^{--}\rangle_{12}\otimes\vert\chi^{--}\rangle_{34}\otimes e^{i\pi/4}U(-iI)\vert\psi\rangle_5
\label{Finalresult2}
\end{equation}
\begin{equation}
2\left({\mathcal P}_-Q_{+}\right)_{34}\vert\Psi\rangle =
{\bf S}\vert \chi^{-+}\rangle_{12}\otimes\vert\chi^{-+}\rangle_{34}\otimes e^{i\pi/4}U(-Y)\vert\psi\rangle_5
\label{Finalresult3}
\end{equation}
\begin{equation}
2\left({\mathcal P}_+Q_{-}\right)_{34}\vert\Psi\rangle =
{\bf S}\vert \chi^{+-}\rangle_{12}\otimes\vert\chi^{+-}\rangle_{34}\otimes e^{i\pi/4}UX\vert\psi\rangle_5
\label{Finalresult4}
\end{equation}
where
\begin{equation}
    {\mathcal P}_\pm =\frac{1}{2}\left({\bf 1}\pm IIXYI\right),\qquad
{\mathcal Q}_\pm =\frac{1}{2}\left({\bf 1}\pm IIZXI\right).
\label{calprojections}
\end{equation}
Now one can notice that
\begin{equation}
     R:=e^{i\pi/4}U=\frac{1}{2}\left[I+i(X+Y+Z)\right]=e^{{i\frac{\alpha}{2}}{\bf n}{\boldsymbol{\sigma}}}\in SU(2),
\end{equation}
where $\alpha=2\pi/3$ and ${\bf n}=\frac{1}{\sqrt{3}}(1,1,1)$.
Hence $R$ is a rotation by $120$ degrees around the axis $\bf n$.

With this notation after omitting the qubit indices we obtain the final result for the decomposition of our state
\begin{equation}
\label{Branching}
\begin{split}
2\vert\Psi\rangle=
&{\bf S} \vert\chi^{+-}\rangle\otimes\vert\chi^{+-}\rangle\otimes RX\vert\psi\rangle-
{\bf S}\vert\chi^{-+}\rangle\otimes\vert\chi^{-+}\rangle\otimes RY\vert\psi\rangle +
\\
&{\bf S}\vert\chi^{++}\rangle\otimes\vert\chi^{++}\rangle\otimes RZ\vert\psi\rangle-
{\bf S}\vert\chi^{--}\rangle\otimes\vert\chi^{--}\rangle\otimes iR\vert\psi\rangle.
\end{split}
\end{equation}
Then from this decomposition one can see that if parties $345$ cooperate then after $34$ performing their joint measurement in the Bell basis $\vert\chi^{\pm\pm}\rangle$, 
with probability $\frac{1}{4}$ they obtain any of the four possible states. After their communicating this result by classical means to party $5$ she/he can first apply the rotation $R$ on the fifth qubit, then depending on the result she/he also applies any of the following unitary operations:$\{X,-Y,Z,-iI\}$ the latter one is just the phase gate $e^{-i\pi/2}$. 
As a result the secret $\vert\psi\rangle$ is recovered for party $5$.

Clearly for the remaining negative line of the doily one has to swap the qubits 
and use the triples (${\color {red}ZX},{\color {red}XY},{\color {red} YZ}$), ($IXZ,IYX,IZY$).
Now the same manipulations have to be done, this time by parties $45$. Then this time the secret is recovered by party $3$.
Moreover, since there is nothing special in the split $(12)(345)$, due to the permutation symmetry of $\vert\Psi\rangle$  one can proceed with any $2+3$  
split of the five qubit system. Hence any party can recover the secret through a joint performance with a couple belonging to a triple she/he is devoted to cooperate with.

\section{The Heptagon Code}

\subsection{The code and its rank-three symplectic geometry}

In the following we will refer to the Calderbank-Shor-Steane (CSS) code\cite{CalderbankShor,Steane} based on the classical Hamming code as the heptagon code\cite{Holohepta} or heptacode in short.
As given in Ref.\cite{NielsenChuang}
its stabilizer group ${\mathcal G}$
in its generator presentation is of the form
\begin{equation}
{\mathcal G}=\langle g_1,g_2,g_3,g_4,g_5,g_6\rangle \equiv\langle
{\color{red}II}I{\color{red}X}XXX,
{\color{red}IX}X{\color{red}I}IXX,
{\color{red}XI}X{\color{red}I}XIX,
{\color{red}II}I{\color{red}Z}ZZZ,
{\color{red}IZ}Z{\color{red}I}IZZ,
{\color{red}ZI}Z{\color{red}I}ZIZ
\rangle.
\label{expl7}
\end{equation}
Now in this notation we have a $3+4$ split. The red digits are occupying the first, the second and the fourth slots.
This choice is dictated by the connection of this quantum code with the classical Hamming code\cite{NielsenChuang}. As it is well-known the Hamming code encodes $4$ classical bits into $7$ ones. The bits $124$ are the check digits and the ones $3567$ are the message ones.
For the rationale of this seemingly strange labeling of these digits we refer to the literature.
We merely note that the numbers $1,2,4$
are the quadratic residues modulo $7$
playing an important role in the constructions of cyclic codes. 
On the other hand the quantum heptacode encodes $1$ logical qubit into $7$ qubits.
As it is also well-known this code also realizes a (4,7) quantum secret sharing threshold scheme\cite{How} our main concern here.
In holograpy this code is playing an important role in generalizing the tensor network construction of the HAPPY code\cite{Happy} based on the pentacode to the heptacode.
It is important to realize that unlike the pentacode the heptacode is not fully permutationally invariant.
However, as was shown in Ref.\cite{Holohepta}
the cyclic symmetry is enought for this code to provide a viable generalization
of the HAPPY code.

As in the case of the pentacode the secret $\vert\psi\rangle=\alpha\vert 0\rangle +\beta\vert 1\rangle$ is encoded into the state
$\vert\Psi\rangle=\alpha\vert\bar{0}\rangle+\beta\vert\bar{1}\rangle$ where
\begin{equation}
    \vert\bar{0}\rangle =NP_1P_2P_3P_4P_5P_6\vert 0000000\rangle,\qquad
    \vert\bar{1}\rangle =XXXXXXX\vert\bar{0}\rangle,
    \nonumber
    \end{equation}
where
\begin{equation}
    P_j=\frac{1}{2}({\bf 1}+g_j).
    \nonumber
\end{equation}
Explicitly one has
\begin{equation}
\begin{split}
\vert\bar{0}\rangle =\frac{1}{\sqrt{8}}(
\vert 0000000\rangle &+
\vert 1010101\rangle +
\vert 0110011\rangle +
\vert 1100110\rangle +\\
\vert 0001111\rangle &+
\vert 1011010\rangle +
\vert 0111100\rangle +
\vert 1101001\rangle),
\end{split}
\label{psi2}
\end{equation}

\begin{equation}
\begin{split}
\vert\bar{1}\rangle =\frac{1}{\sqrt{8}}(
\vert 1111111\rangle &+
\vert 0101010\rangle +
\vert 1001100\rangle +
\vert 0011001\rangle +\\
\vert 0001111\rangle &+
\vert 0100101\rangle +
\vert 1000011\rangle +
\vert 0010110\rangle).
\end{split}
\label{psi3}
\end{equation}
The full set of nontrivial observables featuring the Abelian group $\mathcal G$
is listed in Appendix \ref{sec:app2}.
Here we notice that unlike in the pentagone code these observables are also featuring $42$ observables containing negative signs. 
On the other hand similar to the pentacode our $3+4$ split {\it all} of the $63$ three-qubit observables are (depicted in red)  showing up.
The four-qubit observables (black) are again displaying some interesting finite geometric pattern.

\begin{figure}
\centering
\includegraphics[width=0.90\linewidth]{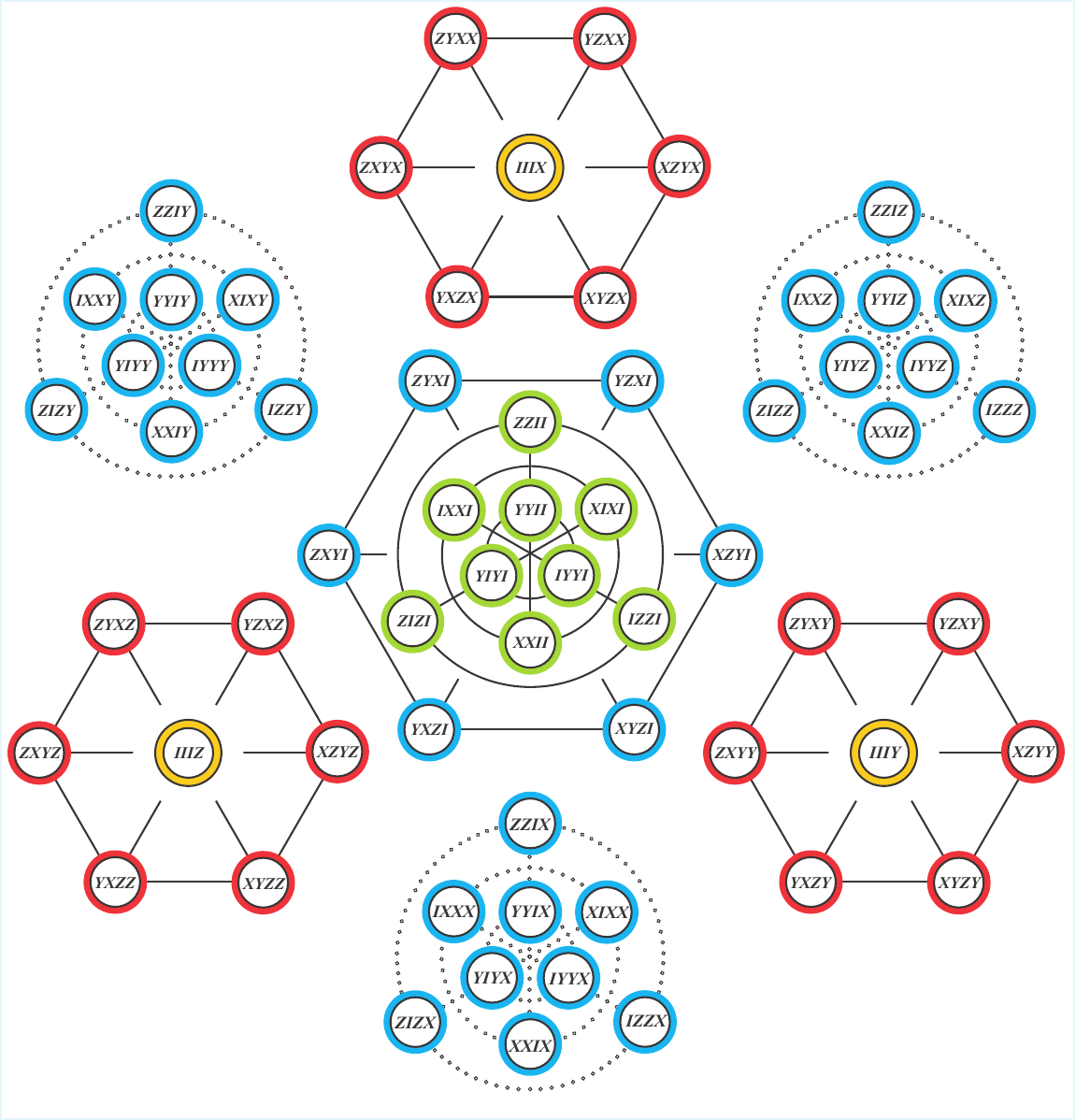}
\caption{\label{heptaly}
The four qubit part of the seven qubit heptacode (CSS-code). In this figure $63$ nontrivial four qubit observables are shown. These are the ones of the heptacode that are showing up in qubits $3567$.
Compare these observables with the relevant (black) part of the full list of $63$ seven qubit ones (after neglecting signs) as given in Appendix \ref{sec:app2}.
In this arrangement the red circles correspond to the $18=9+9+9$ observables not containing the identity. The blue, green and yellow circles contain $33=9+9+9+6$, $9$, $3$ observables.
They are containing the identity in one, two and three slots respectively.
These data identify our embedded configuration
as type $23$ in the classification as given in Table 5. of Ref.\cite{SanigaTax}.
}
\label{fig:heptaly}
\end{figure}

In finite geometric terms 
such structures can be understood in terms of symplectic polar spaces: $W(2N-1,2)$.
For the mathematical background see Appendix \ref{sec:app3}.
In our representation of these abstract mathematical objects $N$ is the number of qubits.
The three-qubit observables ($N=3$) are representing a symplectic polar space $W(5,2)$. This object is having $63$ points $315$ lines and $135$ generators (Fano planes). (For a pictorial representation of three such planes intersecting in a line see Figure~\ref{fig:triple}.)
On the other hand the four-qubit observables ($N=4$) are representing a $W(7,2)$
with $255$ points, $5355$ lines, $11475$
planes and $2295$ generators.
Now the four qubit part of the heptacode
is featuring merely $63$ observables from the possible $255$ ones. Hence the four-qubit part of the heptacode represents an embedding of $W(5,2)$ into
$W(7,2)$.
The classification of $W(5,2)$s living in $W(7,2)$ has been given in Ref.\cite{SanigaTax}.
According to this classification the possible $W(5,2)$s are organized into $29$ types. Except for type $23$, types 21-29 have some extra characteristics discussed in Table 5. of Ref.\cite{SanigaTax}.
Now as an interesting result of this paper Figure~\ref{fig:heptaly} reveals that the characteristics of 
entry $23$ of that Table is precisely encapsulated in the finite geometric structure of the four-qubit part of the heptacode. Some more interesting coincidences with this structure will follow next.

\subsection{Positive planes as stabilizers}

For the pentagon code we have seen that the space of contexts coincides with the set of lines of $W(3,2)$ aka the doily. The set of such $15$ lines coincides with the set of isotropic lines of the projective geometry $PG(3,2)$.
We stressed that these lines can be positive or negative and
we have realized that a positive isotropic line is just
and Abelian group of commuting observables not containing $-{\bf 1}$.

In order to understand quantum secret sharing in a finite geometric terms using the heptacode one has to consider also positive {\it planes}. Since we are over the two element field $GF(2)$ these are Fano planes containing seven points and seven lines see Appendix \ref{sec:app2}. Naively one expects that a positive plane should be defined as a one for which the product of the representative observables is $+{\bf 1}$. 
However,  objects of that kind can still have negative lines.

For example
let us consider a Fano plane which is represented by the following set of $7$ commuting observables also featuring the four qubit part of the $3+4$ splitted version of the heptacode as can be seen in \ref{sec:app2}
\begin{equation}
    (IXXI,IYYI,IZZI,-IXXZ,-IYYZ,IZZZ, IIIZ).
    \label{negplane}
\end{equation}
The product of these observables is $+{\bf 1}:=+IIII$, however
there are also negative lines. For example the line  $(IXXI,IYYI,IZZI)$ is of that kind.

One can remedy this situation by flipping the signs
of some observables.
We call a plane positive if the product of all of its observables is $+{\bf 1}$ and moreover all of its lines are also positive.
Othervise we call a plane negative.
As an example our negative plane can be turned into a positive one by chosing the signs
as follows
\begin{equation}
    (IXXI,-IYYI,IZZI,IXXZ,-IYYZ, IZZZ, IIIZ).
\label{set}
\end{equation}
One can then notice that the set (\ref{set}) taken together with the identity $IIII$ can be elevated to the status of a stabilizer group ${\mathcal P}^{+++}$
with a possible generator presentation
\begin{equation}
    {\mathcal P^{+++}}:=\langle IXXI,IZZI,IIIZ\rangle.
\end{equation}
Now this is a genuine stabilizer subgroup of the four qubit Pauli group with the two dimensional stabilized subspace
\begin{equation}
    \vert\Pi^{+++}\rangle:=(\alpha\vert 0\rangle+\beta\vert 1\rangle)\otimes\frac{1}{\sqrt{2}}(\vert 00\rangle +\vert 11\rangle)\otimes \vert 0\rangle= (\alpha\vert 0\rangle+\beta\vert 1\rangle)\otimes \vert\varphi^{++}\rangle\otimes\vert 0\rangle.
\end{equation}
Clearly one has eight possibilities for obtaining such stabilizer subgroups. They are of the form coming from the eight different arrangement of signs in front of the generators: 
\begin{equation}
    {\mathcal P^{\pm\pm\pm}}:=\langle \pm IXXI,\pm IZZI,\pm IIIZ\rangle.
\label{plane8}
\end{equation}
The $8$  two dimensional stabilized subspaces are of the form
\begin{equation}
    \vert\Pi^{\pm\pm +}\rangle:=(a\vert 0\rangle+b\vert 1\rangle)\otimes \vert\varphi^{\pm\pm}\rangle\otimes\vert 0\rangle
,\qquad
\vert\Pi^{\pm\pm-}\rangle:=(a\vert 0\rangle+b\vert 1\rangle)\otimes \vert\varphi^{\pm\pm}\rangle\otimes\vert 1\rangle.
\label{pistab}
\end{equation}
where $a,b\in{\mathbb C}$ and for our conventions of the Bell states see Eqs.(\ref{Bell1})-(\ref{Bell4}) in Appendix \ref{sec:app1}.

\begin{figure}[htbp]
    \centering
    \includegraphics[width=0.5\textwidth]{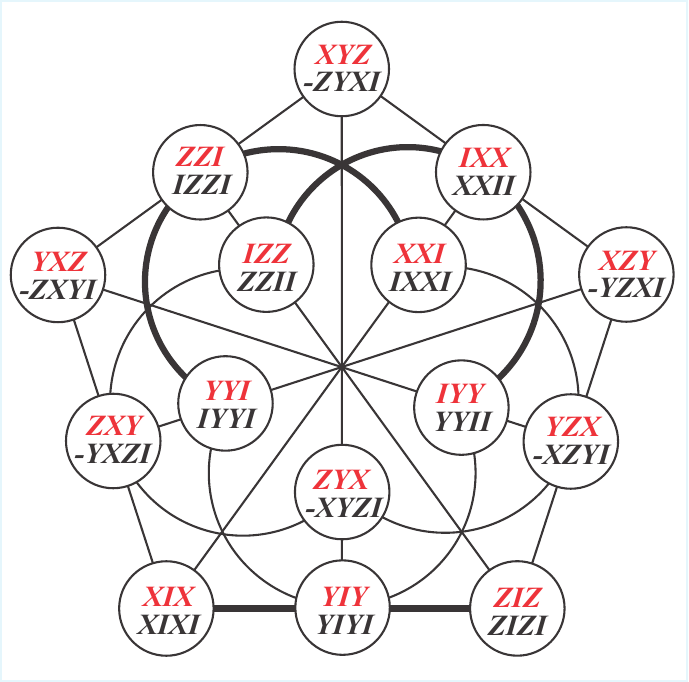}
    \caption{\label{fig:troily} The $15$ observables from the central part of Figure~\ref{fig:heptaly} form a doily i.e. a $W(3,2)$ embedded into $W(7,2)$ shown in black. The red three-qubit labels corresponding to the remaining digits of the $15$ stabilizer group elements form an embedding of a $W(3,2)$ into $W(5,2)$.
    The bold faced lines are negative lines in both cases. The remaining ones are positive. Compare this figure with
Figure~\ref{fig:doily}.
}
    \label{troily}
\end{figure}

\subsection{Breaking the secret with negative planes}
\label{sec:break}
Let us now see how we can implement a protocol for breaking the $(4,7)$ secret sharing scheme using negative planes of the $W(5,2)$ living inside $W(7,2)$.
Let us first choose the negative plane
of Eq.(\ref{negplane}).
One can elevate this plane to a positive one in eight different ways.
These are the planes ${\mathcal P}^{\pm\pm\pm}$ of Eq.(\ref{plane8}).
By an abuse of notation (do not confuse these with the ones of Eqs.(\ref{confuse})) we consider the projectors 
\begin{equation}
P_{\pm}=\frac{1}{2}({\bf 1}\pm {\color{red}II}I{\color{red}I}IIZ),    \label{pe}
\end{equation}
\begin{equation}
Q_{\pm}=\frac{1}{2}({\bf 1}\pm {\color{red}II}I{\color{red}I}XXI),   
\label{qu}
\end{equation}
\begin{equation}
R_{\pm}=\frac{1}{2}({\bf 1}\pm {\color{red}II}I{\color{red}I}ZZI).    \label{er}
\end{equation}
Then one has the following results
\begin{equation}
2Q_+R_+\vert\Psi\rangle =\vert\varphi^{++}\rangle_{12}\otimes \vert\psi\rangle_3\otimes \vert \varphi^{++}\rangle_{47}\otimes \vert\varphi^{++}\rangle_{56}, 
\label{res1}
\end{equation}
\begin{equation}
2Q_-R_+\vert\Psi\rangle =\vert\varphi^{-+}\rangle_{12}\otimes Z\vert\psi\rangle_3\otimes \vert \varphi^{-+}\rangle_{47}\otimes \vert\varphi^{-+}\rangle_{56}, 
\label{res2}
\end{equation}
\begin{equation}
2Q_+R_-\vert\Psi\rangle =\vert\varphi^{+-}\rangle_{12}\otimes X\vert\psi\rangle_3\otimes \vert \varphi^{+-}\rangle_{47}\otimes \vert\varphi^{+-}\rangle_{56}, 
\label{res3}
\end{equation}
\begin{equation}
2Q_-R_-\vert\Psi\rangle =\vert\varphi^{--}\rangle_{12}\otimes XZ\vert\psi\rangle_3\otimes \vert \varphi^{--}\rangle_{47}\otimes \vert\varphi^{--}\rangle_{56}. 
\label{res4}
\end{equation}

Now let us notice that
\begin{equation}
    \sqrt{2}P_+\{\vert\varphi^{++}\rangle,
\{\vert\varphi^{-+}\rangle,
\vert\varphi^{+-}\rangle,
\{\vert\varphi^{-}\rangle\}_{47}=\{\vert 00\rangle,\vert 00\rangle,\vert 10\rangle,-\vert 10\rangle\}_{47},
\nonumber
\end{equation}
\begin{equation}
    \sqrt{2}P_-\{\vert\varphi^{++}\rangle,
\{\vert\varphi^{-+}\rangle,
\vert\varphi^{+-}\rangle,
\{\vert\varphi^{-}\rangle\}_{47}=\{\vert 11\rangle,-\vert 11\rangle,\vert 01\rangle,\vert 01\rangle\}_{47}.
\nonumber
\end{equation}
Then acting with this on our previous result we obtain $8$
terms of the form $\sqrt{8}P_{\pm}Q_{\pm}R_{\pm}\vert\Psi\rangle$.
The sum of these terms gives $\sqrt{8}\vert\Psi\rangle$.
Hence after introducing the notation
\begin{equation}
    \vert\Phi^{\pm\pm 0}\rangle_{jkl} :=\vert\varphi^{\pm\pm}\rangle_{jk}\otimes \vert 0\rangle_l,\qquad
    \vert\Phi^{\pm\pm 1}\rangle_{jkl} :=\vert\varphi^{\pm\pm}\rangle_{jk}\otimes\vert 1\rangle_l
\label{bisepbasis1}
\end{equation}
we obtain the decomposition
\begin{equation}
\begin{split}
    \sqrt{8}\vert\Psi\rangle&=\vert\Phi^{+-0}\rangle_{124}\otimes X\vert\psi\rangle_3\otimes\vert\Phi^{+-1}\rangle_{567}+
\vert\Phi^{+-1}\rangle_{124}\otimes X\vert\psi\rangle_3\otimes\vert\Phi^{+-0}\rangle_{567}\\    
&+
\vert\Phi^{--0}\rangle_{124}\otimes XZ\vert\psi\rangle_3\otimes\vert\Phi^{--1}\rangle_{567}-
\vert\Phi^{--1}\rangle_{124}\otimes XZ\vert\psi\rangle_3\otimes\vert\Phi^{--0}\rangle_{567}\\
&+
\vert\Phi^{++0}\rangle_{124}\otimes \vert\psi\rangle_3\otimes\vert\Phi^{++0}\rangle_{567}+
\vert\Phi^{++1}\rangle_{124}\otimes \vert\psi\rangle_3\otimes\vert\Phi^{++1}\rangle_{567}\\
&+
\vert\Phi^{-+0}\rangle_{124}\otimes Z\vert\psi\rangle_3\otimes\vert\Phi^{-+0}\rangle_{567}-
\vert\Phi^{-+1}\rangle_{124}\otimes Z\vert\psi\rangle_3\otimes\vert\Phi^{-+1}\rangle_{567}.
\end{split}
\label{decomphepta}
\end{equation}

This decomposition shows that if parties $3567$
aware of (\ref{decomphepta}) agree to cooperate,
then after parties $567$ performing a joint measurement in the basis of Eq.(\ref{bisepbasis1}) with probability $1/8$
they obtain the result corresponding to the respective part of the decomposition.
Then parties $567$ communicate their result by classical means to party $3$.
Then depending on this information on the outcome of their measurement, she/he performs either of the unitary operations $I,X,XZ,Z$.
As a result of these manipulations party $3$ will possess the secret.

\begin{figure}[htbp]
    \centering
    \includegraphics[width=0.6\textwidth]{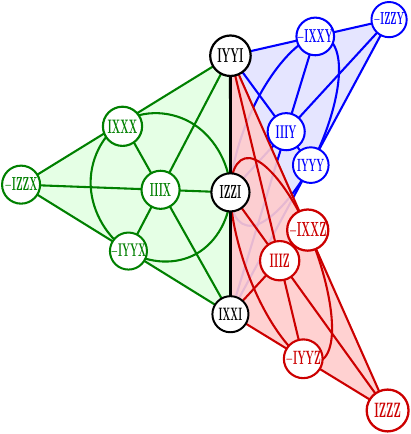}
    \caption{\label{fig:triple} The $15$ observables from the four qubit part of the $3+4$ split of the heptacode containing the identity in the first slot of the four qubit part. This is qubit No:3 in the seven qubit picture. See Appendix \ref{sec:app2}.  These observables can be  arranged into a triple of negative Fano planes intesecting in a line. The Fano planes are negative because they contain negative lines. In particular the black line of intersection is a negative one. The red plane is the one of Eq.(\ref{negplane}) featuring our detailed example of a secret breaking protocol effected by joint manipulations of parties $3567$ of Section \ref{sec:break}. There the secret is finally showing up in qubit No:$3$. As explained in the text there are $8$ different ways for making this triple to contain only {\it positive planes}, i.e. eligible for a stabilizer subgroup 
interpretation for all members of the triple.
    These $8$ different ways correspond to the $8$ terms in the decomposition of Eq.(\ref{decomphepta}).
Our figure should be imagined as a one connected to the corresponding black negative line of the doily of 
Figure~\ref{fig:troily}.
Due to cyclic symmetry in qubits $356$ actually there are {\it three such triples of Fano planes}. The new triples are obtained by cyclic permutation of the observables in the first three slots i.e. in slots $356$ of our figure. The new triples then also should be connected to the relevant black negative doily lines of Figure~\ref{fig:troily}. See also Appendix \ref{sec:heptaplanes}
}
    \label{triple}
\end{figure}

At this point one can observe that actually we have {\it three} different possibilities of negative planes giving rise to three different recovery protocols.
Only one of them is the one associated with the stabilizers ${\cal P}^{\pm\pm\pm}$ of Eq.(\ref{plane8}) giving rise to our protocol based on the (\ref{decomphepta}) decomposition.
In order to reveal the presence of these extra planes the only thing to do is instead of $IIIZ$ to adjoin either the observable $IIIX$
or $IIIY$ to the usual ones $IXXI,IZZI$.
These give rise to two more eight-tuples of stabilizer groups namely
\begin{equation}
    \langle \pm IXXI,\pm IZZI,\pm IIIX\rangle,\qquad
\langle \pm IXXI,\pm IZZI,\pm IIIY\rangle.
\nonumber    
\end{equation}
Hence taken together with Eq.(\ref{plane8}) we have altogether three eight-tuples. They are three eight-tuples of positive planes having their origin in the existence of a triple of negative planes of the four-qubit part of the heptacode. An interesting property of these negative planes is that they are intersecting in a line: the negative line $(IXXI,IYYI,IZZI)$
needed for the derivation of the result encapsulated in Eqs.(\ref{res2})-(\ref{res4}).
Moreover, our triple of planes is based on the three observables $IIIX,IIIY,IIIZ$ depicted by yellow circles in Figure~\ref{fig:heptaly}.
This intersecting configuration of planes is illustrated in Figure~\ref{fig:triple}. Here the triple of planes is illustrated with three colors. The red plane of Figure~\ref{fig:triple} is the one related to our protocol of Eq.(\ref{decomphepta}).
For further discussion on the geometry of these planes and the alternative recoveries associated to them  see Appendix 
\ref{sec:heptaplanes}.

Finally one should realize that one can permute qubits $356$ in a cyclic manner to obtain protocols for recoveries of the secret in the corresponding slots.
Clearly such manipulations are based on two other sets of $15$ four-qubit observables containing an identity in slots $5$ and $6$. Then altogether one has $9$ negative planes, containing triples of intersecting ones in doily lines. For more details on this point see Appendix \ref{sec:heptaplanes}.

\section{Conclusions}
\label{sec:conc}

\subsection{Summary of results}

In this paper we have shown how quantum secret sharing\cite{How} based on the pentagon and heptagon codes can be understood in finite geometric
terms.
We have also demonstrated how in this perspective the secret breaking protocols are connected to the quantum notions of entanglement and contextuality.
We have chosen the pentagon and heptagon codes because they play a pivotal role in building up holographic codes\cite{Happy,Holohepta}.
Then the results of our paper, conceived in the spirit of our recent one\cite{LevayToy},  can also be considered as  ones providing further evidence for the fact that finite geometric stuctures showing up in error correcting codes can be regarded as building blocks for  patterns of emergent spacetime structures.
Here we indeed managed to show that even in the simplest scenarios provided by these codes the basic ingredients of discretized geometry emerge hand in hand with quantum notions like entanglement, contextuality\cite{Bell,KochenSpecker,Mermin}, quantum nets\cite{GHW}
and MUBs\cite{MUB}.

The elaborations of our paper were based on the well-known stabilizer formalism\cite{Gottesman,Calderbank,CalderbankShor}. Our simple strategy was to look at the explicit structure of the $15$ and $63$ nontrivial
observables forming the Abelian stabilizer groups of the pentagon and heptagon codes. From the finite
geometric point of view these observables can be regarded as representatives of points belonging to $3$
and $5$ dimensional totally isotropic subspaces of the projective spaces $PG(9, 2)$ and $PG(13,2)$ equipped
with a natural symplectic form. Taking this symplecic structure into account one can realize that our
codes are
 rather living in the symplectic polar spaces $W(9,2)$ and $W(13,2)$.

Then for the pentagon code we invoked a $2 + 3$ and for the heptagone code a $3 + 4$ split respectively.
Under this split we obtained two families of $15$ and $63$ observables. Unlike the original observables of
the totally isotropic subspaces that are mutually commuting the split families of observables contain
both commuting and not commuting elements. Due to this fact these two families can be used to
parametrize the nontrivial incidence geometries of the symplectic polar spaces $W(3,2)$ (doily) and
$W(5,2)$.

The basic object underlying our elaborations concerning secret breaking protocols then was the space of contexts.
The points of this space were maximal totally isotropic subspaces
of the projective spaces $PG(2n-1,2)$ or equivalently subspaces of dimension $n-1$ (generators) of $W(2n-1,2)$.
Here $n$ was the number of parties who were {\it not} willing to cooperate for breaking the secret.
In particular
for $n=2$ (pentacode)  these points of context space corresponded to (totally) isotropic lines and for $n=3$ (heptacode) totally isotropic planes.
It is also known that
subspaces of dimension $n-1$ of $PG(2n-1,2)$ 
form the Grassmannians ${\mathcal G}(2n-1,n-1)$. See Appendix \ref{sec:app3}.
On the other hand {\it totally isotropic} subspaces of $PG(2n-1,2)$  form the {\it Lagrangian Grassmannian}: ${\mathcal L}{\mathcal G}(2n-1,n-1)$. 
In the cases studied here these spaces are comprising $15$ and $135$ points. Then the points of our spaces of context  corresponded to the $15$ lines of $W(3,2)$ (doily) and $135$ (Fano) planes of $W(5,2)$.

In the $n=2$ case $3$ negative lines of $W(3,2)$ and in the $n=3$ case $9$ negative planes of $W(5,2)$ played an important role.
This role was intimately connected to our $2+3$ split of the pentacode and $3+4$ split of the heptacode. Indeed, due to this split  there also appeared $k=3$ and $k=4$ parties who were willing to cooperate for breaking the secret.
Then as a result of this splitting and the fact that $k>n$ this scenario provided nontrivial embeddings of the spaces $W(2n-1,2)$ into $W(2k-1,2)$.
Then in our considerations lines labeled with three-qubit  and planes labeled with four-qubit observables also appeared.
This upgraded our considerations with the presence of special {\it embedded negative lines} ($k=3$), and {\it special embedded negative planes} ($k=4$).

By elevating these subspaces to positive ones in different ways made us possible to associate to these special points of context space different branches of {\it biseparable states}. The presence of these biseparable branches in the decomposition of the pentagon and heptagon stabilizer states $\vert\Psi\rangle$ of the (\ref{psi}), (\ref{psi2})-(\ref{psi3}) form  made it possible for one of the parties to possess the secret $\vert\psi\rangle$ of Eq.(\ref{secret}).
We have given the different decompositions
explicitly. They are displaying the protocols needed to be performed by the cooperating parties enabling one of them to succeed in achieving her/his goal.
For the explicit decompositions see Eqs.(\ref{branching}) and (\ref{Branching}) for the pentacode and
Eqs.(\ref{decomphepta}), (\ref{decompbluehepta}), (\ref{decompgreenhepta}) for the heptacode.
These decompositions also show the basic role of entangled Bell-states playing in performing a secret breaking protocol.

During our considerations we have also realized that the four-qubit part of the heptacode represents a very special embedding of $W(5,2)$ into
$W(7,2)$.
Indeed, according to the classification of $W(5,2)$s living in $W(7,2)$ of Ref.\cite{SanigaTax} the possible $W(5,2)$s are organized into $29$ types. Except for type $23$, the types $21-29$ of Table 5. of Ref.\cite{SanigaTax} are showing some extra characteristics. 
The elaborations of this paper revealed that the missing characteristic feature of 
entry $23$ of that Table can be identified
as the one displaying the finite geometric structure of the four-qubit part of the heptacode. Compare
Figure~\ref{fig:heptaly} and Table 5. of Ref.\cite{SanigaTax} in this respect.
Some other interesting finite geometric coincidences also follow.
In this respect note that in our considerations the doily i.e. $W(3,2)$ played a basic role. It is then interesting to notice that the symplectic polar space $W(9,2)$ hosting our pentacode, is the
smallest rank space
that contains doilies featuring $12$, i.e. the maximum possible number of
negative lines\cite{Muller}.
On the other hand $PG(13,2)$, the ambient space of $W(13,2)$ where
our heptacode lives, is also the space where the split
Cayley hexagon of order two\cite{GenPol} can be embedded universally\cite{Thas}, i.e. where
each of its $16383$ geometric hyperplanes
originates from its intersection
with some of $16383$ $PG(12,2)$s of $PG(13,2)$.
It is an interesing coincidence since the split Cayley hexagon  has already made its debut to physics in connection with black hole solutions in string theory in Refs.\cite{Hexa,Planat}.
To cap all this, these solutions has also been connected\cite{upto} to the structure of the classical Hamming code which is forming the basis for constructing our heptacode.
These connections deserve further scrutiny. 

\subsection{A comment on a space-time interpretation of context space}

\label{sec:comment}

Let us finally elaborate on our results from the perspective of a dual representation
of our space of contexts.
This representation is provided by the Plücker map\cite{Penrose,Notable,LevayToy}.
This is a map from $PG(2n-1,2)$ to a subset of $PG(N-1,2)$ where 
$N=\binom{2n}{n}$. Under this map different geometric objects on one side correspond to geometric objects on the other.
Moreover, thanks to this correspondence the symplectic form on $PG(2n-1,2)$ defines a symplectic form on $PG(N-1,2)$ too. 
Considering the polarizations of these forms, see Eq.(\ref{kvad}), one also has quadratic forms on both sides of the correspondence.

The $n=2$ and $N=6$ case gives rise to a $GF(2)$ version\footnote{In twistor theory one is working not over the two element field $GF(2)={\mathbb Z}_2$ but rather over the field of complex numbers $\mathbb C$. Here instead of a projective space like $PG(3,2):=PG(3,{\mathbb Z}_2)$ we have $PG(3,{\mathbb C}):={\mathbb C}{\mathbb P}^3$ which is called the projective twistor space\cite{Penrose}. Under the twistor program the geometry of compactified and complexified Minkowski space is reformulated in terms of the complex geometry of ${\mathbb C}{\mathbb P}^3$.} of the famous Klein correspondence forming the basis of Roger Penrose's twistor program\cite{Penrose}. (For a detailed illustration of this point and the analogy with twistor theory see Ref.\cite{LevayToy}.) Under this geometric objects of $PG(3,2)$ are in correspondence with objects belonging to a hyperbolic quadric $Q^+(5,2)$ (called the Klein quadric) which is in $PG(5,2)$.
In particular the $35$ lines of $PG(3,2)$ correspond to
the $35$ points of $Q^+(5,2)$.
What is interesting for us is that under this correspondence the $15$ {\it isotropic lines} of $PG(3,2)$, aka the generators of $W(3,2)$,
correspond to $15$ special points in $Q^+(5,2)$. 

Due to the fact that we have also a quadratic form at our disposal on $Q^+(5,2)$ one can show that the Klein quadric is the $GF(2)$ analogue of conformally compactified and complexified Minkowski space time\cite{Penrose,LevayToy}. Moreover, the quadratic form is playing the role of the one originating from the well-known Minkowski inner product, with a caveat that in the discrete case we are over $GF(2)$ then for two points in $Q^+(5,2)$ we have only either light-like or non light-like separation.

What is interesting for us is that under the Klein correspondence the $15$ {\it isotropic lines} of $PG(3,2)$, aka the generators of $W(3,2)$,
correspond to $15$ special points in $Q^+(5,2)$. 
Then one can prove\cite{LevayToy} that the $15$ points of our space of contexts are forming the "real" points\footnote{Of course unlike inside the field $\mathbb C$ inside $GF(2)$ it makes no sense to talk about the field of the real numbers $\mathbb R$. However, it makes sense to enforce a constraint which is very similar to the reality condition of twistor theory\cite{LevayToy}.} of the Klein quadric i.e forming the discrete $GF(2)$ analogue of compactified ordinary Minkowski space-time.
Moreover, two lines in $W(3,2)$ are intersecting if and only if the corresponding points are light-like separated.
In this curious space-time geometry a light ray consists of only three-points: these points form a line of the "real part" of the Klein quadric.
Then the surprising result is that this real part is forming another copy of a doily.
Hence we have the nice result:  under the Klein correspondence the space of contexts comprising the lines of a doily in $PG(3,2)$ is represented by the points of another doily this time living inside $Q^+(5,2)$ as a $GF(2)$ version of Minkowski space.
Moreover, under the correspondence intersecting contexts correspond to light-like separated points\cite{LevayToy}.

Now in the case of the pentagon code we have our
$\vert\Psi\rangle$ and our space of contexts consisting of $15$ points in the Klein quadric forming a discretized $GF(2)$ version of Minkowski space at our disposal.
Moreover, each context is associated with a decomposition of $\vert\Psi\rangle$. In particular the two contexts from the nonintersecting three negative ones studied in Section \ref{sec:break}
give rise to "space-time" points that are not light-like separated.
These two ones
give rise to the decompositions of Eq.(\ref{branching}) and
Eq.(\ref{Branching}). They are featuring four branches.
Both decompositions are employing biseparable states, hence eligible for secret breaking.
Moreover, applying the swap operation one can even include a third non-collinear space-time point (corresponding to the third negative line of the doily) with the corresponding third eligible decomposition for secret breaking. 

Clearly for all of the remaining $13$ points (representing $13$ other contexts) there can also be defined decompositions with four branches, but they are not featuring biseparable states, hence not eligible ones for secret breaking. Moreover, they are not necessarily featuring only two qubit entangled Bell states but can also contain three-qubit GHZ ones and two qubit separable ones\footnote{It is known that stabilizer states from the remaining  $W$ class cannot show up.}.
In any case a consistent assignment to our space-time points elements of orthogonal branches can only exist locally.
They are like local sections of a bundle over discretized compactified Minkowski space. According to our argument presented in Appendix \ref{sec:appdoily}
there is no global section to this bundle of frames. In this sense our discretized Minkowski space (the doily again) is contextual.

Now a very interesting observation is in order.
The situation encoded into the pentacode gives an amusing example of 
a five-qubit
Universe which is in the quantum state $\vert\Psi\rangle$.
Since all of its observables are mutually commuting this universe by itself is devoid of any interesting geometric (causal) structure. However, after employing our $2+3$ split the doily structure is emerging in both partitions. Moreover, after the split negative lines are also showing up.
Now, we have just explained that the doily is geometrically the $GF(2)$ version of compactified real Minkowski spacetime.
We then have just found a nice toy model for the emergence of compactified Minkowski spacetime geometry for observers constrained to have access only to a part of the  Universe.

Via local sections based on spreads capable of covering all observables on the two qubit part of the split, one can perform measurements and explore some parts of the geometry of this Minkowski space.
On the other hand since this space is contextual there are no global sections, then there are necessarily incompatible sections corresponding to incompatible measurements obstructing the exploration of the global structure of our geometry.
However, there exist local sections which are providing quantum nets familiar from Ref.\cite{GHW}.
See also Section \ref{sec:appdoily} in this respect. These are directly related to MUBs\cite{MUB}.
Hence in this discretized spacetime picture one can clearly see how different quantum notions are interconnected with the geometric structure of an emergent spacetime.

Clearly one can generalize these results even for the heptacode. This time the Plücker map defines a correspondence\cite{Planat,LevayToy} between the $135$ contexts of $PG(5,2)$ and the $135$ points of a hyperbolic quadric $Q^+(7,2)$. However, now instead of a quadratic form we have a so called chronometric form living on an object called quantum space-time familiar from the work of Brody and Hughston\cite{Brody}.
This object keeps track of the more sophisticated intersection properties of contexts manifesting in the more complicated separation properties of quantum space time points.
In any case biseparable decompositions of $\vert\Psi\rangle$ assigned to quantum space time points still take care of the branches responsible for consistent secret breaking protocols. 
However, at the time of writing this paper we do not know whether the corresponding bundle is nontrivial or trivial.
In order to settle this issue we have to see (possibly by a numerical calculation) whether $W(5,2)$ is plane contextual (with plane contextuality as defined in this paper) or not. 

One can also realize that our method of splits can  be used for CSS\cite{CalderbankShor} codes for establishing important connections between quantum notions like entanglement and contextuality and finite geometry.
In this case one can exploit the generalized map existing\cite{LevayToy} between context space inside $PG(2n-1, 2)$ and the points of $Q^+(2^n-1,2)$. We hope that we have convinced the reader that by employing finite geometric techniques one can obtain further interesting results of geometric nature similar to the ones derived in this paper.
In particular one can study other threshold schemes of quantum secret sharing in the hope to arrive at a fully general finite geometric picture.
Such issues we are intending to study in future work.

In the book of Thomas Hertog\cite{Hertog} Chapter 6. has the title: "No Question? No History!"
This title refers to the top down approach initiated by Stephen Hawking and the author for understanding the 
history of the Universe whose structure is inherently connected to questions asked by observers.
In our finite geometric models,  split scenarios 
correspond to harassing the system with special questions.
Our results show that under this process nontrivial finite geometries show up.
Hence, paraphrasing Hawking and Hertog we can say:
"No Question? No Geometry!"

\section{Acknowledgement}
 
This work was supported by the HUN-REN Hungarian Research Network through the Supported Research Groups Programme, HUN-REN-BME-BCE Quantum Technology Research Group (TKCS-2024/34) as well as by the Slovak VEGA grant agency, project number 2/0043/2. We are also grateful to our friends
Petr Pracna and Zsolt Szabó for their technical assistance with preparing figures.

\section{Appendix}

\subsection{Some simple stabilized subspaces}
\label{sec:app1}

In this subsection we work out the structures associated with the triples ${\color {red} XX}YIY,{\color {red} YY}ZIZ,{\color {red} ZZ}XIX\in {\mathcal S}$ and
${\color {red} YX}XYI,{\color {red} ZY}YZI,{\color {red} XZ}ZXI\in {\mathcal S}$.
As we have discussed in the main text after applying the $2+3$ split the resulting triples of two (${\color {red}XX},{\color {red}YY},{\color {red} ZZ}$),
(${\color {red}YX},{\color {red}ZY},{\color {red} XZ}$
)  and the corresponding three ($YIY,ZIZ,XIX$)
, ($XYI,YZI,ZXI$)
qubit observables yield negative lines of the corresponding
doilies (See Figure~\ref{fig:doily}.)
Stabilizer groups cannot contain minus the identity\cite{NielsenChuang} hence we can only elevate these triples to the status of a stabilizer group by flipping some appropriate number of signs.

For the two qubit case these are the four groups featuring  observables acting only on the first two qubits of our five qubit Hilbert space:

\begin{equation}
 {\mathcal S}^{++}:=\langle XX,ZZ\rangle=\{II,XX,-YY,ZZ\}, 
\nonumber
\end{equation}

\begin{equation}
 {\mathcal S}^{+-}:=\langle XX,-ZZ\rangle=\{II,XX,YY,-ZZ\},
\nonumber
\end{equation}

\begin{equation}
 {\mathcal S}^{-+}:=\langle -XX,ZZ\rangle=\{II,-XX,YY,ZZ\},
\nonumber
\end{equation}

\begin{equation}
 {\mathcal S}^{--}:=\langle -XX,-ZZ\rangle=\{II,-XX,-YY,-ZZ\}.
\nonumber
\end{equation}
These four stabilizer groups
determine (up to a phase) four stabilized states: the four Bell states
\begin{equation}
\vert\varphi^{++}\rangle
=\frac{1}{\sqrt{2}}(\vert 00\rangle +\vert 11\rangle),
\label{Bell1}
\end{equation}

\begin{equation}
\vert\varphi^{+-}\rangle
=\frac{1}{\sqrt{2}}(\vert 01\rangle +\vert 10\rangle),
\label{Bell2}
\end{equation}

\begin{equation}
\vert\varphi^{-+}\rangle
=\frac{1}{\sqrt{2}}(\vert 00\rangle -\vert 11\rangle),
\label{Bell3}
\end{equation}

\begin{equation}
\vert\varphi^{--}\rangle
=\frac{1}{\sqrt{2}}(\vert 01\rangle -\vert 10\rangle).
\label{Bell4}
\end{equation}

For the three qubit part we have the corresponding stabilizer groups

\begin{equation}
 {S}^{++}:=\langle YIY,XIX\rangle=\{III,YIY,XIX,-ZIZ\}, 
\label{pp}
\end{equation}

\begin{equation}
 {S}^{+-}:=\langle YIY,-XIX\rangle=\{III,YIY,-XIX,ZIZ\},
\label{pm}
\end{equation}

\begin{equation}
 {S}^{-+}:=\langle -YIY,XIX\rangle=\{III,-YIY,XIX,ZIZ\},
\label{mp}
\end{equation}

\begin{equation}
 {S}^{--}:=\langle -YIY,-XIX\rangle=\{III,-YIY,-XIX,-ZIZ\}.
\label{mm}
\end{equation}
These groups stabilize two dimensional subspaces of the form
\begin{equation}
    \vert\Phi^{++}\rangle=\frac{1}{\sqrt{2}}(\vert 01\rangle +\vert 10\rangle)_{35}\otimes (a\vert 0\rangle +b\vert 1\rangle )_4,
    \label{subs1}
\end{equation}
\begin{equation}
    \vert\Phi^{+-}\rangle=\frac{1}{\sqrt{2}}(\vert 00\rangle -\vert 11\rangle)_{35}\otimes (a\vert 0\rangle +b\vert 1\rangle )_4,
    \label{fipm}
\end{equation}
\begin{equation}
    \vert\Phi^{-+}\rangle=\frac{1}{\sqrt{2}}(\vert 00\rangle +\vert 11\rangle)_{35}\otimes (a\vert 0\rangle +b\vert 1\rangle )_4,
\end{equation}
\begin{equation}
\vert\Phi^{--}\rangle=\frac{1}{\sqrt{2}}(\vert 01\rangle -\vert 10\rangle)_{35}\otimes (a\vert 0\rangle +b\vert 1\rangle )_4,
\label{subs4}
\end{equation}
with $a,b\in{\mathbb C}$.
According to the SLOCC classification (stochastic local operations assisted with classical communication) of three-qubit states\cite{Dur} these are belonging to one of the  three biseparable classes where the fourth qubit is separated from the rest.

In the main text we also use the stabilized states associated to the triple
$(XY,YZ,ZX)$. 
This and its swapped version shows up in the triple
${\color {red} YX}XYI,{\color {red} ZY}YZI,{\color {red} XZ}ZXI\in {\mathcal S}$.
The corresponding four stabilizer groups
are
\begin{equation}
 {\mathcal G}^{++}:=\langle XY,ZX\rangle=\{II,XY,-YZ,ZX\}, 
\nonumber
\end{equation}

\begin{equation}
 {\mathcal G}^{+-}:=\langle XY,-ZX\rangle=\{II,XY,YZ,-ZX\},
\nonumber
\end{equation}

\begin{equation}
 {\mathcal G}^{-+}:=\langle -XY,ZX\rangle=\{II,-XY,YZ,ZX\},
\nonumber
\end{equation}

\begin{equation}
 {\mathcal G}^{--}:=\langle -XY,-ZX\rangle=\{II,-XY,-YZ,-ZX\}.
\nonumber
\end{equation}

These four stabilizer groups
determine four stabilized states similar to the usual four Bell states
\begin{equation}
\vert\chi^{++}\rangle
=\frac{1}{2}\left(\vert 00\rangle +\vert 01\rangle -i \vert 10\rangle +i\vert 11\rangle\right)
=\frac{1}{\sqrt{2}}(\vert \tilde{1}0\rangle +\vert \tilde{0}1\rangle),
\label{bell1}
\end{equation}

\begin{equation}
\vert\chi^{--}\rangle=
\frac{1}{2}\left(\vert 00\rangle -\vert 01\rangle -i \vert 10\rangle -i\vert 11\rangle\right)
=\frac{1}{\sqrt{2}}(\vert \tilde{1}0\rangle -\vert \tilde{0}1\rangle),
\label{bell2}
\end{equation}

\begin{equation}
\vert\chi^{-+}\rangle=
\frac{1}{2}\left(\vert 00\rangle +\vert 01\rangle +i \vert 10\rangle -i\vert 11\rangle\right)
=\frac{1}{\sqrt{2}}(\vert \tilde{0}0\rangle +\vert \tilde{1}1\rangle),
\label{bell3}
\end{equation}

\begin{equation}
\vert\chi^{+-}\rangle=
\frac{1}{2}\left(\vert 00\rangle -\vert 01\rangle +i \vert 10\rangle +i\vert 11\rangle\right)
=\frac{1}{\sqrt{2}}(\vert \tilde{0}0\rangle -\vert \tilde{1}1\rangle).
\label{bell4}
\end{equation}
Here for the eigenstates of the observable $Y$ we introduced the notation
\begin{equation}
\vert\tilde{0}\rangle =\frac{1}{\sqrt{2}}\left(\vert 0\rangle +i\vert 1\rangle\right),\qquad
\vert\tilde{1}\rangle =\frac{1}{\sqrt{2}}\left(\vert 0\rangle -i\vert 1\rangle\right).
\label{basetilde}
\end{equation}
These new basis states define the unitary
\begin{equation}
    U=\frac{1}{\sqrt{2}}\begin{pmatrix}
    1&1\\i&-i\end{pmatrix}
\label{Yunitary}
\end{equation}
permuting the basic observables as
\begin{equation}
    U^{\dagger}XU=Y,\qquad
    U^{\dagger}YU=Z,\qquad
    U^{\dagger}ZU=X.
\end{equation}

\subsection{Contextuality and the Doily}
\label{sec:appdoily}
Have a look again at our doily of Figure~\ref{fig:doily}.
Its set of totally isotropic lines of cardinality $15$ represents the maximal sets of mutually commuting observables defined up to sign. We will conform with Refs. \cite{Planat,degree} and call it {\it the space of contexts}.
We have seen that one can {\it always} tackle the sign ambiguities associated to this space {\it locally}. This means that at most $12$ lines can be elevated {\it consistently} to the status of a genuine stabilizer subgroup of the two qubit Pauli group.
Elevated here means that one can associate to a context (an isotropic line) one of the four possible stabilizer subgroups.
Or in other words this means to associate to a context one of its four possible stabilizer states.
However, one cannot do this {\it globally}.
The minimal number of contexts obstructing this lifting procedure from the space of contexts to the  stabilizers is called the {\it degree of contextuality} of the finite geometric configuration.
In the case of the doily it is $3$.
Notice that this definition is not the original one as given for example in Ref.\cite{degree} but more like in the spirit of Ref.\cite{Khu} 
and references therein.

The geometry of the situation is reminiscent of a discrete version of fiber bundles.
A trivial bundle is a one that can be written as a direct product of two spaces, the base space and the fiber. The canonical example in this respect is the direct product of a circle (base) and an interval (fiber). The resulting space is the total space of the bundle. In our example it is a cylinder. A nontrivial fiber bundle is a one which {\it locally} looks like the direct product of these spaces.
However, globally it is not, it is rather a twisted collection of local trivializations. For our example: a twisted product of the circle and the interval can yield the Möbius band.

In our discrete case of the doily our base space is the space of contexts, and the fiber is formed by quadruplets of orthogonal elements taken from the four dimensional complex vector space of two qubits.
Then the total space of the bundle is formed by special sets of quadruplets taken from the set of orthonormal frames.
According to Ref.\cite{Khu} and references therein the right language to talk about such structures and their generalizations is rather sheaf theory not the theory of bundles.
However, here our arguments are intuitive hence we will use the bundle picture as a guiding principle and refrain from attempting to formalize a precise mathematical setup.

A section of a bundle is a map from the base space to the total space. It is like taking a slice of the bundle.  
Now it is known that a bundle is trivial if one can find a {\it global section} of it. 
In our example of the cylinder produced as the direct product of a circle and the interval, global sections correspond to different slicings of the cylinder.
The bundle is nontrivial if one can only find {\it local sections}.
Then in our intuitive picture the two qubit doily ($W(3,2)$) is contextual because the associated bundle is nontrivial.
On the other hand we call a configuration  {\it noncontextual} if one can associate a trivial bundle to it.
An example of that kind is the $15$ element set of triples formed by the codewords of the pentacode.
Indeed, taken together the red plus black labeling of Fig~\ref{fig:doily} all the lines comprising triples formed from  pentads of observables are positive proving the triviality.
Another way of saying this is that the corresponding totally isotropic subspace of $PG(9,2)$ is line noncontextual\cite{SanigaTax}.
Then in this bundle picture via our $2+3$ split from a trivial bundle (a noncontextual configuration) one can generate two nontrivial ones (two contextual configurations).
Hence by our approach the physical picture of splitting a  system (characterized by a highly entangled stabilizer state of the pentacode)  into two subsystems is represented by this geometric idea of creating nontrivial bundles from a trivial one.
A geometric idea which is inherently connected to the important notions of entanglement and contextuality.

Since the doily is contextual its associated bundle admits at most local sections.
Now there exist local sections of a very important kind.
They are featuring only $5$ contexts corresponding to disjoint lines that partition the point set of the doily. We will call this set of $5$ elements a {\it patch} of the space of contexts.
An example of that kind is the partition
\begin{equation}
    \{ ZZ,IZ,ZI\},\{IY,YI,YY\},\{XX,IX,XI\},\{XY,ZX,YZ\},\{ZY,XZ,YX\}.
\end{equation}
Since the last two lines are negative, one can elevate them to positive ones any way we please by flipping some signs accordingly. Moreover, one can even play this game with positive lines in a way preserving their positivity. For example
\begin{equation}
    \{ ZZ,IZ,ZI\},\{IY,-YI,-YY\},\{XX,IX,XI\},\{XY,-ZX,YZ\},\{ZY,-XZ,YX\}.
    \label{confi}
\end{equation}
This provides a local section of our bundle.
The peculiarity of this local section is its rotational symmetry. This means that there exists a unitary ${\mathcal U}\in U(4)$ transformation with the property ${\mathcal U}^5={\bf 1}$ which permutes the elements of the triples cyclically under conjugation by 
keeping the ordering inside the triples
\cite{LevayToy,Wootters}.
Hence for example we have
$\mathcal{U}(IX){\mathcal U}^{\dagger}=-ZX$,
$\mathcal{U}(-ZX){\mathcal U}^{\dagger}=-XZ$ etc.
This means that starting from the stabilizer state $\vert 00\rangle$ associated to the positive line $\{ZZ,IZ,ZI\}$, by simply applying $\mathcal U$ all the other four stabilizer states can be obtained. Hence this local section can be generated from a single point belonging to the patch. 
This process generalizes nicely\footnote{However, according to our knowledge, the possibility of this nice bundle interpretation have not been noticed yet.} to an arbitrary number of qubits and is known in the literature under the name "associating a quantum net to the points of a discrete phase  space"\cite{GHW,Wootters,LevayToy}.
Moreover, starting with the elements of the canonical quadruplet of orthonormal vectors  corresponding to the stabilizers $\langle \pm IZ,\pm ZI\rangle$, the quadruplets of sections generated in this rotationally covariant way give rise to an association of a MUB, i.e. a mutually unbiased basis system\cite{MUB} to this patch of the space of contexts of the doily\cite{GHW,LevayToy}.

It turns out that there are {\it six} possible partitions of the point set of the doily. These partitions are called spreads of the doily\cite{MUBfinite}. (Alternatively they are {\it isotropic} spreads of $PG(3,2)$.)
For a list of such partitions see Table II. of Ref.\cite{LevayToy}.
Then one can cover the space of contexts with six "coordinate patches" each of them consisting of
the five "points" of the corresponding spread. Some of these patches are intersecting others are not. One can then define local sections as above and look how the intersecting ones (different MUBs) are related to each other.
What we find that now these local sections cannot be patched together consistently to form a global section due to the contextuality of the underlying configuration\cite{LevayToy}.

Notice also that if we regard the (\ref{confi}) configuration as the $2$-qubit part of our $2+3$ split then the pentagon code also associates to it a corresponding $3$ qubit one of the form
\begin{equation}
 \{XIX,YYZ,ZYY\},
\{XXY,-YXX,-ZIZ\},
 \{YIY,ZZY,XZZ\},
 \{IYX,-IXZ,IZY\},
 \{YZI,-ZXI,XYI\}.
 \nonumber
\end{equation}
This defines another local section. This time it associates to the line contexts two dimensional subspaces.
We also discover two of our negative lines at the last two entries.
The corresponding subspaces have been used for our secret breaking protocols.
As it was crucial for the secret breaking these subspaces were spanned by biseparable states. 
In particular the sign distribution in the last triple of our local section corresponds to one possible outcome of the secret breaking protocol. 
In our special case this outcome is the one associated with the projectors
${\mathcal P}_+$ and ${\mathcal Q}_-$ of Eq.(\ref{calprojections}) that project to the first term of the (\ref{Branching}) decomposition of $\vert\Psi\rangle$.
Hence our local section combined with the $2+3$ split of the pentacode encodes a particular branch
of outcomes of a secret breaking protocol.
Clearly the other branches associated with other local sections work in the same manner.

What about the other members of our spread?
In this respect one can show that the first three entries are featuring linear combinations of three-qubit states belonging to the GHZ class of Ref.\cite{Dur}.
They will provide other branches with other types of decompositions for $\vert\Psi\rangle$.
However, since they are {\it not} of the biseparable type needed for a {single party} to possess finally the secret, they are not eligible for implementing a secret breaking scheme.

\subsection{The full set of nontrivial observables of the heptacode}
\label{sec:app2}
In this explicit list of the nontrivial observables the product of generators is represented by writing the corresponding numbers in increasing order from left to right eg: $1246:=g_1g_2g_4g_6$.
The $g_j, j=1,2,3,4,5,6$ generators of the code are given by the set of Eq.(\ref{expl7}).
Let us use the notation $\overline{14}=2356$, $\overline 4=12356$ etc.
Then the $63$ nontrivial observables of the Abelian group $\mathcal G$ are

\begin{equation}
\begin{split}
    (1,2,&3,4,5,6)=\\
&({\color{red}II}I{\color{red}X}XXX,
{\color{red}IX}X{\color{red}I}IXX,
{\color{red}XI}X{\color{red}I}XIX,
{\color{red}II}I{\color{red}Z}ZZZ,
{\color{red}IZ}Z{\color{red}I}IZZ,
{\color{red}ZI}Z{\color{red}I}ZIZ),
\end{split}
\nonumber
\end{equation}

\begin{equation}
\begin{split}    (\overline{1},&\overline{2},\overline{3},\overline{4},\overline{5},\overline{6})=\\
&(-{\color{red}YY}I{\color{red}Z}XXZ,
-{\color{red}YZ}X{\color{red}Y}IXZ,
-{\color{red}ZY}X{\color{red}Y}XIZ,
-{\color{red}YY}I{\color{red}X}ZZX,
-{\color{red}YX}Z{\color{red}Y}IZX,
-{\color{red}XY}Z{\color{red}Y}ZIX),
\end{split}
\nonumber
\end{equation}

\begin{equation}
    \begin{pmatrix}12&13&14&15&16&\\&23&24&25&26\\&&34&35&36\\&&&45&46\\&&&&56\end{pmatrix}=
\begin{pmatrix}
{\color {red}IX}X{\color {red} X}XII&
{\color {red}XI}X{\color {red} X}IXI&
{\color {red}II}I{\color {red} Y}YYY&
-{\color {red}IZ}Z{\color {red} X}XYY&
-{\color {red}ZI}Z{\color {red} X}YXY\\
&{\color {red}XX}I{\color {red} I}XXI&
-{\color {red}IX}X{\color {red} Z}ZYY&
{\color {red}IY}Y{\color {red} I}IYY&
-{\color {red}ZX}Y{\color {red} I}ZXY\\
&&-{\color {red}XI}X{\color {red} Z}YZY&
-{\color {red}XZ}Y{\color {red} I}XZY&
{\color {red}YI}Y{\color {red} I}YIY\\
&&&{\color {red}IZ}Z{\color {red} Z}ZII&
{\color {red}ZI}Z{\color {red} Z}IZI\\
&&&&{\color {red}ZZ}I{\color {red} I}ZZI
\end{pmatrix},
\nonumber
\end{equation}

\begin{equation}
    \begin{pmatrix}\overline{12}&\overline{13}&\overline{14}&\overline{15}&\overline{16}&\\&\overline{23}&\overline{24}&\overline{25}&\overline{26}\\&&\overline{34}&\overline{35}&\overline{36}\\&&&\overline{45}&\overline{46}\\&&&&\overline{56}\end{pmatrix}=
\begin{pmatrix}
-{\color {red}YZ}X{\color {red} Z}XIY&
-{\color {red}ZY}X{\color {red} Z}IXY&
{\color {red}YY}I{\color {red} I}YYI&
-{\color {red}YX}Z{\color {red} Z}XYI&
-{\color {red}XY}Z{\color {red} Z}YXI\\
&-{\color {red}ZZ}I{\color {red} Y}XXY&
-{\color {red}YZ}X{\color {red} X}ZYI&
{\color {red}YI}Y{\color {red} Y}IYI&
-{\color {red}XZ}Y{\color {red} Y}ZXI\\
&&-{\color {red}ZY}X{\color {red} X}YZI&
-{\color {red}ZX}Y{\color {red} Y}XZI&
{\color {red}IY}Y{\color {red} Y}YII\\
&&&-{\color {red}YX}Z{\color {red} X}ZIY&
-{\color {red}XY}Z{\color {red} X}IZY\\
&&&&-{\color {red}XX}I{\color {red} Y}ZZY
\end{pmatrix},
\nonumber
\end{equation}

\begin{equation}
\begin{pmatrix}156&146&145\\
256&246&245\\356&346&345\end{pmatrix}=-
\begin{pmatrix}
{\color {red}ZZ}I{\color {red} X}YYX&
{\color {red}ZI}Z{\color {red} Y}XYX&
{\color {red}IZ}Z{\color {red} Y}YXX&
\\
{\color {red}ZY}X{\color {red} I}ZYX&
{\color {red}ZX}Y{\color {red} Z}IYX&
{\color {red}IY}Y{\color {red} Z}ZXX&
\\
{\color {red}YZ}X{\color {red} I}YZX
&{\color {red}YI}Y{\color {red} Z}XZX&
{\color {red}XZ}Y{\color {red} Z}YIX
\end{pmatrix},
\qquad 123={\color{red} XX}I{\color{red}X}IIX,
\nonumber
\end{equation}

\begin{equation}
\begin{pmatrix}234&134&124\\
235&135&125\\236&136&126\end{pmatrix}=-
\begin{pmatrix}
{\color {red}XX}I{\color {red} Z}YYZ&
{\color {red}XI}X{\color {red} Y}ZYZ&
{\color {red}IX}X{\color {red} Y}YZZ&
\\
{\color {red}XY}Z{\color {red} I}XYZ&
{\color {red}XZ}Y{\color {red} X}IYZ&
{\color {red}IY}Y{\color {red} Z}ZXX&
\\
{\color {red}YX}Z{\color {red} I}YXZ
&{\color {red}YI}Y{\color {red} X}ZXZ&
{\color {red}ZX}Y{\color {red} X}YIZ
\end{pmatrix},
\qquad 456=
{\color{red} ZZ}I{\color{red}Z}IIZ,
\nonumber
\end{equation}

\begin{equation}
    123456={\color{red}YY}I{\color{red}Y}IIY.
    \nonumber
\end{equation}

\subsection{Some geometric properties of the nine negative planes of the heptacode.}

\label{sec:heptaplanes}

In Section \ref{sec:break}
we used the red Fano plane of  Figure~\ref{fig:triple}. for presenting an explicit protocol for breaking the secret. As a result of this the third party managed to recover the secret.
There we mentioned that there are alternative ways for this recovery process for the third party playing a distinguished role.
Moreover, we also stressed that the same types of manipulations apply for the recovery at the third, the fifth and sixth slots as well.
Then altogether there are 
nine Fano planes for such recovery protocols.
Here we give an explicit list of these planes and comment on some of their geometric properties.

For recovery at the third, the fifth and sixth slots one has to look at the sructure of heptacode observables containing the identity operator at the third, the fifth and sixth slots.
Then one can notice that these three sets can be organized into intersecting Fano planes.
Moreover, it is also useful to split the relevant seven qubit observables according to our usual $(124)(3567)$ manner to three-qubit and four-qubit parts. 

Then the list of planes in the $(124)+(3567)$ notation is as follows. For recovery in the third slot one can use either of the following three planes:
\begin{equation}
    \{{\color{blue}YYY},YYI,ZZI,XXI,{\color{blue}ZZY,IIY,XXY}\},\quad \{{\color{blue}IIIY}, IYYI,IZZI,IXXI,{\color{blue}-IXXY,IYYY,-IZZY}\}
\nonumber
\end{equation}

\begin{equation}
    \{{\color{red}ZZZ},YYI,ZZI,XXI,{\color{red}YYZ,XXZ,IIZ}\},\quad \{{\color{red} IIIZ}, IYYI,IZZI,IXXI,{\color{red}-IXXZ,-IYYZ,IZZZ}\}
\nonumber
\end{equation}

\begin{equation}
    \{{\color{green}XXX},YYI,ZZI,XXI,{\color{green}IIX,ZZX,YYX}\},\quad \{{\color{green} IIIX}, IYYI,IZZI,IXXI,{\color{green}IXXX,-IYYX,-IZZX}\}
\nonumber
\end{equation}
\noindent
intersecting in the lines $(XXI,YYI,ZZI)$ and $(IXXI,IYYI,IZZI)$.

For recovery in the fifth slot:
\begin{equation}
    \{{\color{blue}YYY},YIY,ZIZ,XIX,{\color{blue}ZYZ,IYI,XYX}\},\quad \{{\color{blue}IIIY}, YIYI,ZIZI,XIXI,{\color{blue}-XIXY,YIYY,-ZIZY}\}
\nonumber
\end{equation}

\begin{equation}
    \{{\color{red}ZZZ},YIY,ZIZ,XIX,{\color{red}YZY,XZX,IZI}\},\quad \{{\color{red} IIIZ}, YIYI,ZIZI,XIXI,{\color{red}-XIXZ,-YIYZ,ZIZZ}\}
\nonumber
\end{equation}

\begin{equation}
    \{{\color{green}XXX},YIY,ZIZ,XIX,{\color{green}IXI,ZXZ,YXY}\},\quad \{{\color{green} IIIX}, YIYI,ZIZI,XIXI,{\color{green}XIXX,-YIYX,-ZIZX}\}
\nonumber
\end{equation}
\noindent
intersecting in the lines $(XIX,YIY,ZIZ)$ and $(XIXI,YIYI,ZIZI)$.

Finally for the recovery in the sixth slot:
\begin{equation}
    \{{\color{blue}YYY},IYY,IZZ,IXX,{\color{blue}YZZ,YII,YXX}\},\quad \{{\color{blue}IIIY}, YYII,ZZII,XXII,{\color{blue}-XXIY,YYIY,-ZZIY}\}
\nonumber
\end{equation}

\begin{equation}
    \{{\color{red}ZZZ},IYY,IZZ,IXX,{\color{red}ZYY,ZXX,ZII}\},\quad \{{\color{red} IIIZ}, YYII,ZZII,XXII,{\color{red}-XXIZ,-YYIZ,ZZIZ}\}
\nonumber
\end{equation}

\begin{equation}
    \{{\color{green}XXX},IYY,IZZ,IXX,{\color{green}XII,XZZ,XYY}\},\quad \{{\color{green} IIIX}, YYII,ZZII,XXII,{\color{green}XXIX,-YYIX,-ZZIX}\}
\nonumber
\end{equation}
\noindent
intersecting in the lines $(IXX,IYY,IZZ)$ and $(XXII,YYII,ZZII)$.

Notice also that the three blue, red and green planes are intersecting in the points $(YYY)(IIIY)$, $(ZZZ) (IIIZ)$ and $(XXX)(IIIX)$ respectively according to the $(124)(3567)$ split.
Hence the same colored planes are intersecting in points and differently colored planes are intersecting in lines.

In Ref.\cite{Planat} it has been shown that the $135$ planes of $W(5,2)$ can be mapped bijectively to the $135$ points of the hyperbolic quadric $Q^+(7,2)$ in $W(7,2)$ by the Plücker map. For the definition of this quadric see Appendix \ref{sec:app3}. In the same paper an  explicit list of this correspondence between planes and points has also been given.
In physical terms this means that there is a bijection between the space of three-qubit plane contexts of cardinality $135$
and the set of four qubit symmetric observables (i.e. ones that contain only an even number of $Y$ observables) of the same cardinality\cite{Planat}.
It has been shown that two points in 
$Q^+(7,2)$ are collinear if and only if the corresponding planes in $W(5,2)$ are intersecting.
A consequence of this result is that the nine planes listed above can be mapped to nine points of this quadric connected by six lines. This means that they form a grid, i.e. a Mermin square\cite{Mermin} in 
$Q^+(7,2)$.
Using the results of Ref.\cite{Planat} one can even calculate the Plücker coordinates of this map and determine the $9$ symmetric four-qubit observables
corresponding to our nine planes. A calculation then shows that

\begin{equation}
    \{{\color{blue}YYY},YYI,ZZI,XXI,{\color{blue}ZZY,IIY,XXY}\}\leftrightarrow {\color{blue}IYYI}
\nonumber    
\end{equation}

\begin{equation}
    \{{\color{red}ZZZ},YYI,ZZI,XXI,{\color{red}YYZ,XXZ,IIZ}\}\leftrightarrow {\color{red}IXXI}
\nonumber    
\end{equation}

\begin{equation}
    \{{\color{green}XXX},YYI,ZZI,XXI,{\color{green}IIX,ZZX,YYX}\}\leftrightarrow {\color{green}IZZI}
\nonumber
\end{equation}

\begin{equation}
    \{{\color{blue}YYY},YIY,ZIZ,XIX,{\color{blue}ZYZ,IYI,XYX}\}\leftrightarrow {\color{blue}IYIY}
\nonumber
\end{equation}

\begin{equation}
    \{{\color{red}ZZZ},YIY,ZIZ,XIX,{\color{red}YZY,XZX,IZI}\}\leftrightarrow {\color{red}IXIX}
\nonumber
\end{equation}

\begin{equation}
    \{{\color{green}XXX},YIY,ZIZ,XIX,{\color{green}IXI,ZXZ,YXY}\}\leftrightarrow {\color{green}IZIZ}
\nonumber
\end{equation}
\noindent

\begin{equation}
    \{{\color{blue}YYY},IYY,IZZ,IXX,{\color{blue}YZZ,YII,YXX}\}\leftrightarrow {\color{blue}IIYY}
\nonumber
\end{equation}

\begin{equation}
    \{{\color{red}ZZZ},IYY,IZZ,IXX,{\color{red}ZYY,ZXX,ZII}\}\leftrightarrow {\color{red}IIXX}
\nonumber
\end{equation}

\begin{equation}
    \{{\color{green}XXX},IYY,IZZ,IXX,{\color{green}XII,XZZ,XYY}\}\leftrightarrow {\color{green}IIZZ}
\nonumber
\end{equation}
\noindent
From here one can easily check that these nine points form a grid located on our quadric. There are six lines in this grid.
There are three negative lines corresponding to differently colored points, and three positive lines for points having the same color.

Notice that the first slot of the $Q^+(7,2)$ four-qubit label is not playing any role here.
However, it is also known that from the $135$ points of our quadric there are $63$ ones that form an object called the split Cayley hexagon\cite{GenPol,Planat}. For its physical role played in black hole solutions in string theory see Refs.\cite{Hexa,upto}.
It turns out that the $63$ special points are corresponding to those symmetric four-qubit observables that are commuting with the special observable $YIII$. Clearly all of our $9$ grid observables are of this type. Hence these points are on the hexagon. However, it is also known that only the {\it negative} lines of our grid in the quadric form also hexagon lines\cite{Planat}.
Finally notice that the same grid also shows up in the 
four-qubit labels of the central doily of
Figure~\ref{fig:troily} with the dummy identity this time located in the last slot.

\subsection{Alternative secret breaking by other negative planes}

In Section \ref{sec:break}
we have seen how to break the secret using the red negative plane of Figure~\ref{fig:triple}.
Here we present protocols based on the blue and green planes of this Figure.

The starting point is the set of Eqs.(\ref{res1})-(\ref{res4})
featuring the projectors $Q_{\pm}$ and $R_{\pm}$ of Eqs.(\ref{qu})-(\ref{er}).
In order to proceed instead of the projectors $P_{\pm}$ of (\ref{pe}) we introduce two other ones
\begin{equation}
U_{\pm}=\frac{1}{2}({\bf 1}\pm {\color{red}II}I{\color{red}I}IIY),   \label{u}
\end{equation}
\begin{equation}
V_{\pm}=\frac{1}{2}({\bf 1}\pm {\color{red}II}I{\color{red}I}IIX).    \label{v}
\end{equation}

For the protocol associated with the {\it blue plane} of Figure~\ref{fig:triple} we perform the simple calculation
\begin{equation}
    \sqrt{2}U_+\{\vert\varphi^{++}\rangle,
\vert\varphi^{-+}\rangle,
\vert\varphi^{+-}\rangle,
\vert\varphi^{--}\rangle\}_{47}=\{\vert \tilde{1}\tilde{0}\rangle,\vert \tilde{0}\tilde{0}\rangle,-i\vert \tilde{0}\tilde{0}\rangle,-i\vert \tilde{1}\tilde{0}\rangle\}_{47},
\nonumber
\end{equation}
\begin{equation}
    \sqrt{2}U_-\{\vert\varphi^{++}\rangle,
\vert\varphi^{-+}\rangle,
\vert\varphi^{+-}\rangle,\vert\varphi^{--}\rangle\}_{47}=\{\vert \tilde{0}\tilde{1}\rangle,\vert \tilde{1}\tilde{1}\rangle,\vert \tilde{1}\tilde{1}\rangle, \vert \tilde{0}\tilde{1}\rangle\}_{47}.
\nonumber
\end{equation}
Then one can calculate the   eight $\sqrt{8}U_{\pm}Q_{\pm}R_{\pm}\vert\Psi\rangle$ states, the sum of them giving
$\sqrt{8}\vert\Psi\rangle$. Collecting everything we obtain

\begin{equation}
\begin{split}
    \sqrt{8}\vert\Psi\rangle&=\vert\Sigma^{+-0}\rangle_{124}\otimes ZY\vert\psi\rangle_3\otimes\vert\Sigma^{+-0}\rangle_{567}+
\vert\Sigma^{+-1}\rangle_{124}\otimes YZ\vert\psi\rangle_3\otimes\vert\Sigma^{+-1}\rangle_{567}\\    
&+
\vert\Sigma^{--0}\rangle_{124}\otimes Y\vert\psi\rangle_3\otimes\vert\Sigma^{--1}\rangle_{567}-
\vert\Sigma^{--1}\rangle_{124}\otimes Y\vert\psi\rangle_3\otimes\vert\Sigma^{--0}\rangle_{567}\\
&+
\vert\Sigma^{++0}\rangle_{124}\otimes \vert\psi\rangle_3\otimes\vert\Sigma^{++1}\rangle_{567}+
\vert\Sigma^{++1}\rangle_{124}\otimes \vert\psi\rangle_3\otimes\vert\Sigma^{++0}\rangle_{567}\\
&+
\vert\Sigma^{-+0}\rangle_{124}\otimes Z\vert\psi\rangle_3\otimes\vert\Sigma^{-+0}\rangle_{567}-
\vert\Sigma^{-+1}\rangle_{124}\otimes Z\vert\psi\rangle_3\otimes\vert\Sigma^{-+1}\rangle_{567},
\end{split}
\label{decompbluehepta}
\end{equation}
where

\begin{equation}
    \vert\Sigma^{\pm\pm 0}\rangle_{jkl} :=\vert\varphi^{\pm\pm}\rangle_{jk}\otimes \vert \tilde{0}\rangle_l,\qquad
    \vert\Sigma^{\pm\pm 1}\rangle_{jkl} :=\vert\varphi^{\pm\pm}\rangle_{jk}\otimes\vert \tilde{1}\rangle_l,
\label{bisepbasis2}
\end{equation}
where for the definitions in (\ref{bisepbasis2}) we use Eq.(\ref{basetilde}).
Hence for the blue plane parties $567$ perform measurements in the biseparable basis of (\ref{bisepbasis2}). Then after communicationg their result to party $3$ he/she performs the usual operations featuring Pauli gates to recover the secret.

For the protocol associated with the {\it green plane} of Figure~\ref{fig:triple} the alternative calculation needed is the one
\begin{equation}
    \sqrt{2}V_+\{\vert\varphi^{++}\rangle,
\vert\varphi^{-+}\rangle,
\vert\varphi^{+-}\rangle,
\vert\varphi^{--}\rangle\}_{47}=\{\vert \hat{0}\hat{0}\rangle,\vert \hat{1}\hat{0}\rangle,\vert \hat{0}\hat{0}\rangle,\vert \hat{1}\hat{0}\rangle\}_{47},
\nonumber
\end{equation}
\begin{equation}
    \sqrt{2}V_-\{\vert\varphi^{++}\rangle,
\vert\varphi^{-+}\rangle,
\vert\varphi^{+-}\rangle,
\vert\varphi^{--}\rangle\}_{47}=\{\vert \hat{1}\hat{1}\rangle,\vert \hat{0}\hat{1}\rangle,-\vert \hat{1}\hat{1}\rangle, -\vert \hat{0}\hat{1}\rangle\}_{47},
\nonumber
\end{equation}
where
\begin{equation}
  \vert\hat{0}\rangle =\frac{1}{\sqrt{2}}(\vert 0\rangle +\vert 1\rangle),\qquad
\vert\hat{1}\rangle =\frac{1}{\sqrt{2}}(\vert 0\rangle -\vert 1\rangle).
\end{equation}
Define now the biseparable states as before
\begin{equation}
    \vert\Omega^{\pm\pm 0}\rangle_{jkl} :=\vert\varphi^{\pm\pm}\rangle_{jk}\otimes \vert \hat{0}\rangle_l,\qquad
    \vert\Omega^{\pm\pm 1}\rangle_{jkl} :=\vert\varphi^{\pm\pm}\rangle_{jk}\otimes\vert \hat{1}\rangle_l.
\label{bisepbasis3}
\end{equation}
Then similar manipulations yield the alternative decomposition
\begin{equation}
\begin{split}
    \sqrt{8}\vert\Psi\rangle&=\vert\Omega^{+-0}\rangle_{124}\otimes X\vert\psi\rangle_3\otimes\vert\Omega^{+-0}\rangle_{567}-
\vert\Omega^{+-1}\rangle_{124}\otimes X\vert\psi\rangle_3\otimes\vert\Omega^{+-1}\rangle_{567}\\    
&+
\vert\Omega^{--0}\rangle_{124}\otimes ZX\vert\psi\rangle_3\otimes\vert\Omega^{--1}\rangle_{567}+
\vert\Omega^{--1}\rangle_{124}\otimes XZ\vert\psi\rangle_3\otimes\vert\Omega^{--0}\rangle_{567}\\
&+
\vert\Omega^{++0}\rangle_{124}\otimes \vert\psi\rangle_3\otimes\vert\Omega^{++0}\rangle_{567}+
\vert\Omega^{++1}\rangle_{124}\otimes \vert\psi\rangle_3\otimes\vert\Omega^{++1}\rangle_{567}\\
&+
\vert\Omega^{-+0}\rangle_{124}\otimes Z\vert\psi\rangle_3\otimes\vert\Omega^{-+1}\rangle_{567}+
\vert\Omega^{-+1}\rangle_{124}\otimes Z\vert\psi\rangle_3\otimes\vert\Omega^{-+1}\rangle_{567}.
\end{split}
\label{decompgreenhepta}
\end{equation}
Then for the green plane parties $567$ perform this time the measurements in the biseparable basis of (\ref{bisepbasis3}). Then after communicating their result to party $3$ she/he performs the usual Pauli gates to recover the secret.

\subsection{Finite geometric structures}
\label{sec:app3}

Here for the convenience of the reader we summarize the finite geometric structures underlying our considerations.

Let  us consider the $2n$-dimensional vector space $V\equiv V(2n,2)$ over the field $GF(2)\equiv \mathbb{Z}_2=\{0,1\}$.
In the following we will refer to the dimension of a vector space as its {\it rank}, hence $V$ is of rank $2n$.
In the canonical basis $e_{\mu}, \mu=0,1,2,\dots 2n-1$ 
we arrange the components of a vector $v\in V$ in the form
\begin{equation}
v\leftrightarrow (q_0,q_1,\dots,q_{n-1},p_0,p_1,\dots,p_{n-1}).
\label{elsokonv}
\end{equation}
\noindent
We represent $n$-qubit observables by vectors of $V$  in the following manner.
Use the mapping
\begin{equation}
(00)\leftrightarrow I, \quad (01)\leftrightarrow X,
\quad (11)\leftrightarrow Y,\quad (10)\leftrightarrow Z,
\label{alap}
\end{equation}
\noindent
where $(X,Y,Z)\equiv (\sigma_x,\sigma_y,\sigma_z)$ i.e they are the usual Pauli spin matrices.
Then for example in the $n=3$ case the array
$(q_0,q_1,q_2,p_0,p_1,p_2)$ of six numbers taken from $\mathbb{Z}_2$ encodes a three-qubit observable {\it up to sign}.
For example the observable $XYZ$ is represented by the vector $(011110)$.
As a result of this procedure we have a map
\begin{equation}
v\mapsto \pm \mathcal{O}_v
\label{op}
\end{equation}
\noindent
between vectors of $V$ and $n$-qubit observables $\mathcal{O}$ up to sign. Note that since under multiplication the operators also pick up multiplicative factors of $\pm i$, generally the space of observables is {\it not} forming an algebra.
In order to properly incorporate the multiplicative structure the right object to consider is the Pauli group\cite{NielsenChuang} which is the set of operators of the form $\{\pm \mathcal{O}_v,\pm i\mathcal{O}_v\}$.  
Then one can show that the center of the Pauli group is the group $\{\pm 1,\pm i\}$, and its central quotient is just the vector space $V$. Under this isomorphism vector addition in $V$ corresponds to multiplication of Pauli group elements up to $\pm 1$ and $\pm i$ times the identity operator. 

The vector space $V$ is also equipped with a symplectic form $\langle\cdot,\cdot\rangle$ encoding the commutation properties of the corresponding observables. 
Namely, for two vectors $v,v^{\prime}\in V$ with components in the canonical basis 
\begin{equation}
v\leftrightarrow (q_0,q_1,\dots,q_{n-1},p_0,p_1,\dots,p_{n-1}),\qquad
v^{\prime}\leftrightarrow (q^{\prime}_0,q^{\prime}_1,\dots,q^{\prime}_{n-1},p^{\prime}_0,p^{\prime}_1,\dots,p^{\prime}_{n-1}),
\label{familiar}
\end{equation}
\noindent
\begin{equation}
\langle v,v^{\prime}\rangle = 
\sum_{i=0}^{n-1}\left(q_ip^{\prime}_i+q^{\prime}_ip_i\right)\in GF(2).
\label{symp}
\end{equation}
\noindent
In the symplectic vector space $(V,\langle\cdot,\cdot\rangle )$ we have $ \langle v,v^{\prime}\rangle=0$ or $1$, referring to the cases when the corresponding $n$-qubit observables are commuting or anticommuting: $[\mathcal{O}_v,\mathcal{O}_{v^{\prime}}]=0$ or
$\{\mathcal{O}_v,\mathcal{O}_{v^{\prime}}\}=0$.
Notice that over the two element field $GF(2)$ an alternating form like $\langle\cdot,\cdot\rangle$ is symmetric.

Since $V$ is even dimensional and the symplectic form is
nondegenerate, the invariance group of the symplectic form is the
symplectic group $Sp(2n,{\mathbb Z}_2)\equiv Sp(2n,2)$. This group is acting on the row
vectors of $V$ via $2n\times 2n$ matrices $M\in Sp(2n,2)$ from the
right, leaving the matrix $J=\langle e_{\mu},e_{\nu}\rangle$ of the symplectic form invariant \begin{equation}
v\mapsto vM,\qquad MJM^t=J. \label{transz} \end{equation} \noindent It is
known that $Sp(2n,2)$ is generated by transvections\cite{Cerchiai}
$T_w\in Sp(2n,2), w\in V$ of the form \begin{equation} T_w:V\to V,\qquad
v\mapsto T_wv=v+\langle v,w\rangle w \label{transvections} \end{equation}
\noindent and they are indeed symplectic, i.e. \begin{equation} \langle
T_wv,T_wv^{\prime}\rangle=\langle v,v^{\prime}\rangle. \label{szimpltulajd} \end{equation}

Given the symplectic form one can define a quadratic form $Q:V\to GF(2)$ by the formula
\begin{equation}
Q(v)\equiv \sum_{i=0}^{n-1}q_ip_i
\label{kvad}
\end{equation}
\noindent
which is related to the symplectic form via 
\begin{equation}
\langle v,v^{\prime}\rangle=Q(v+v^{\prime})+Q(v)+Q(v^{\prime}).
\label{kapcsolat}
\end{equation}
\noindent
Generally quadratic forms which, by a convenient choice of basis, can be given the (\ref{kvad}) canonical form are called {\it hyperbolic}\cite{Cerchiai}.
In our special case the meaning of the (\ref{kvad}) quadratic form is clear: for $n$-qubit observables $\mathcal{O}_v$ containing an even (odd) number of tensor product factors of $Y$, the value of $Q$ is zero (one). Hence the observables which are symmetric under transposition are having $Q(v)=0$ and ones that are antisymmmetric under transposition are having $Q(v)=1$.

In the following we will refer to the set of subspaces of rank $k=1,2,3,\dots ,2n-1$ of $V$ as the Grassmannians: $Gr(k,2n)$.
For $k=1,2,3,\dots,2n-1$ these spaces are arising by considering the spans of one, two, three, etc., $2n-1$  linearly independent vectors $v_1,v_2,v_3,\dots ,v_{2n-1}\in V$. These Grassmannians are just sets of {\it lines}, {\it planes}, {\it spaces}, etc., and {\it hyperplanes} through the origin. Since we are over $GF(2)$ these are sets of the form $\{au\}$, $\{au+bv\}$, $\{au+bv+cw\}$, etc. with $a,b,c\in \mathbb{Z}_2$, consisting of one, three, seven, etc., $2^{2n-1}-1$ nontrivial vectors.
A plane over $GF(2)$ containing seven points and seven lines is a Fano plane.
A line $\{au\}$ through the origin (zero vector) is a {\it ray}. Over $GF(2)$ the number of rays equals the number of nonzero vectors of $V$. 

Regarding the set of rays as the set of points of a new space of one dimension less, gives rise to the projective space ${\mathbb P}(V)\equiv PG(2n-1,2)$. In this projective context lines and planes in $V$
correspond to points and lines of ${\mathbb P}(V)$. Hence the collection of the projectivization of the Grassmannians $G(k,2n)$ 
denoted by $\mathcal{G}(k-1,2n-1)$
forms the projective geometry of $PG(2n-1,2)$. 
Usually  the word {\it dimension} is used for the dimension of a projective subspace ${\mathbb P}(S)$, and {\it rank} is used for the dimension of the corresponding vector subspace $S$.
Notice that according to  Eq.(\ref{op}), the sets of projective subspaces (points, lines, planes, etc. hyperplanes) of $PG(2n-1,2)$, i.e. the Grassmannians $\mathcal{G}(0,2n-1), \mathcal{G}(1,2n-1), \mathcal{G}(2,2n-1),\dots \mathcal{G}(2n-2,2n-1)$, up to a sign correspond to the set of {\it nonzero} observables, certain triples, seven-tuples , etc., $2^{2n-1}-1$-tuples of them.

Since we have a symplectic form at our disposal
one can further specify the projective subspaces and the corresponding observables. A subspace ${\mathcal I}$ of $V$ is called {\it isotropic} if there is a vector in it which is orthogonal to to the whole subspace. 
${\mathcal S}$ is {\it totally isotropic} if for all points $u,v\in V$ 
we have $\langle u, v\rangle =0$ i.e. a totally isotropic subspace is orthogonal to itself.
Notice that in the cases of rank one and two subspaces of $V$ i.e. zero and one projective dimensional subspaces of $P(V)$
(points and lines in the corresponding projective geometry) the notions of isotropic and totally isotropic coincide.
By virtue of Eq.(\ref{kapcsolat}) it is also clear that an ${\mathcal S}$ is totally isotropic if for all $v\in {\mathcal S}$ we have $Q(v)=0$.

It is clear that a totally isotropic subspace is represented by a set of mutually commuting observables. In this formalism a stabilizer group is based on a  totally isotropic subspace of $P(V)$.
For example in the case of the pentagon code we have a totally isotropic subspace of dimension $3$
of the projective geometry
$PG(9,2)$.
Notice that the dimension of maximally totally isotropic subspaces of $PG(2n-1,2)$ is $n-1$.
These are called Lagrangian subspaces of $PG(2n-1,2)$.
These are arising from  totally isotropic rank $n$ subspaces of the rank $2n$ vector space $(V,\langle\cdot,\cdot \rangle)$.
The corresponding set of observables corresponds to a maximal set of $2^{n}-1$ tuples of mutually commuting observables.
The set of maximal totally isotropic subspaces of $PG(2n-1,2)$ is called the Lagrangian Grassmannian
${\mathcal L}{\mathcal G}(n-1,2n-1)$. The number of points of this space\cite{Lagrass} is $\prod_{i=1}^n(2^i+1)$.
In the cases studied in this paper we have for $n=2$ ($n=3$)  $15$ ($135$) points. They are corresponding to the totally isotropic lines (planes) of $PG(3,2)$ ($PG(5,2)$).

The incidence structure of the set of totally isotropic subspaces of $PG(2n-1,2)$ defines the {\it symplectic polar space} of rank $n$:
$\mathcal{W}(2n-1,2)$.
A projective subspace of maximal dimension is called one of its {\it generators}. Clearly the set of generators of $W(2n-1,2)$ forms the Lagrangian Grassmannian ${\mathcal L}{\mathcal G}(n-1,2n-1)$. 
$\mathcal{W}(2n-1,2)$ made its debut to physics in connection with observables of $n$-qubits in Ref.\cite{SP}.
Geometrically the doily playing a basic role of the pentagon code is ${\mathcal W}(3,2)$.

For our quadratic form of Eq.(\ref{kvad}) the points satisfying the equation $Q(v)=0$ form a {\it hyperbolic quadric} in $PG(2n-1,2)$ denoted by $Q^+(2n-1,2)$.
Hence symmetric $n$-qubit observables are represented by points {\it on}, and antisymmetric ones {\it off} this hyperbolic quadric in $PG(2n-1,2)$.
It is then easy to prove that each hyperbolic quadric contains $(2^{n-1}+1)(2^n-1)$ points. The quadric embedded in $PG(5,2)$ under the Plücker map is called the Klein quadric $Q^+(5,2)$. It contains $35$ points.
And indeed, it is easy to check that the number of nontrivial three-qubit symmetric observables (the ones containing an even number of $Y$s) is $35$.
For a detailed description of the Plücker map and the Klein correspondence mentioned in Section \ref{sec:comment} see Ref.\cite{LevayToy}.

\bibliographystyle{alpha}
\bibliography{sample}

\end{document}